\documentclass[12pt,a4paper]{article}
\makeatletter
\def\eqnarray{%
\stepcounter{equation}%
\let\@currentlabel=\theequation
\global\@eqnswtrue
\global\@eqcnt\z@
\tabskip\@centering
\let\\=\@eqncr
$$\halign to \displaywidth\bgroup\@eqnsel\hskip\@centering
$\displaystyle\tabskip\z@{##}$&\global\@eqcnt\@ne
\hfil$\displaystyle{{}##{}}$\hfil
&\global\@eqcnt\tw@$\displaystyle\tabskip\z@{##}$\hfil
\tabskip\@centering&\llap{##}\tabskip\z@\cr}
\makeatother
\newcommand{\kansu}[2]{{{#1}\!\left({#2}\right)}}
\newcommand{\ket}[1]{{\vert{#1}\rangle}}
\newcommand{\bra}[1]{{\langle{#1}\vert}}
\newcommand{\kett}[2]{{\vert{#1,#2}\rangle}}
\newcommand{\calh}{{\cal H}}
\newcommand{\calm}{{\cal M}}
\newcommand{\cala}{{\cal A}}
\newcommand{\calf}{{\cal F}}

\newcommand{\fukuso}{{\mathbf C}}
\newcommand{\real}{{\mathbf R}}
\newcommand{\futon}{{\bf N}}

\newcommand{\st}{{St_2}}
\newcommand{\stm}{{St_m}}
\newcommand{\grm}{{Gr_m}}
\newcommand{\eem}{{E_m}}
\newcommand{\muzeta}{{\vert\mu\vert}}
\newcommand{\nuzeta}{{\vert\nu\vert}}
\newcommand{\xizeta}{{\vert\xi\vert}}
\newcommand{\zezeta}{{\vert\zeta\vert}}
\newcommand{\alzeta}{{\vert\alpha\vert}}
\newcommand{\bezeta}{{\vert\beta\vert}}
\newcommand{\gazeta}{{\vert\gamma\vert}}
\newcommand{\beazeta}{{\vert\beta_1\vert}}
\newcommand{\bebzeta}{{\vert\beta_2\vert}}
\newcommand{\lam}{{\bf \lambda}}
\newcommand{\slam}{{\bf \lambda_0}}

\begin{document}

\title{\sl Mathematical Foundations of Holonomic \\
           Quantum Computer II}
\author{
  Kazuyuki FUJII
  \thanks{E-mail address : fujii@math.yokohama-cu.ac.jp }\\
  Department of Mathematical Sciences\\
  Yokohama City University\\
  Yokohama 236-0027, Japan
  }
\date{}
\maketitle\thispagestyle{empty}
%
%
%
%
\begin{abstract}
  This is a sequel to the papers (quant-ph/9910063) and 
  (quant-ph/0004102). The aim of this paper is to give  
  mathematical foundations to Holonomic Quantum Computation (Computer)
  proposed by Zanardi and Rasetti (quant-ph/9904011) and 
  Pachos and Chountasis (quant-ph/9912093).
 
  In 2-qubit case we give an explicit form to non-abelian Berry connection 
  of quantum computational bundle which is associated with Holonomic
  Quantum Computation, on some parameter space. 

  We also suggest a possibility that not only usual holonomy but also 
  higher-dimensional holonomies must be used to prove a universality of 
  our Holonomic Quantum Computation. 
\end{abstract}

\newpage

%
%
%
%

\section{Introduction}

This is a sequel to the papers \cite{KF2} and \cite{KF3}.

Quantum Computer is a very attractive and challenging task in New 
Millennium.
\par 
\noindent
After the breakthrough by P. Shor \cite{PS} there has been remarkable
progress in Quantum Computer or Computation (QC briefly).
This discovery had a great influence on scientists. This drived not only 
theoreticians to finding other quantum algorithms, but also 
experimentalists to building quantum computers.
See \cite{AS} and \cite{RP} in outline. \cite{LPS} is also very useful.

On the other hand, Gauge Theories are widely recognized as the basis in 
quantum field theories.
Therefore it is very natural to intend to include gauge theories 
in QC $\cdots$ a construction of ``gauge theoretical'' quantum computation
or of ``geometric'' quantum computation in our terminology. 
The merit of geometric method of QC may be strong for the influence 
from the environment. See for example \cite{JVEC}. 

In \cite{ZR} and \cite{PZR} Zanardi and Rasetti proposed an attractive idea 
$\cdots$ Holonomic Quantum Computation (Computer) $\cdots$
using the non-abelian Berry phase (quantum holonomy in the mathematical 
language). See also \cite{AYK} and \cite{JP} as another interesting 
geometric models.
In their model a Hamiltonian (including some parameters) must be
degenerated because an adiabatic connection is introduced using
this degeneracy \cite{SW}. 
In other words, a quantum computational bundle is introduced on some 
parameter space due to this degeneracy (see \cite{ZR}) and the canonical 
connection of this bundle is just the one above.

They gave a few simple but interesting examples to explain their idea. We 
believe that these examples will become important in the near future. 
But their works (\cite{ZR}, \cite{PZR} and \cite{PC}) are a bit coarse in 
the mathematical point of view. Moreover, for more than 2-qubit case 
a universality of Holonomic Quantum Computation has not been proved. 
Therefore in this paper we will attack this problem in the case of 
2-qubit. 

Namely we construct a quantum computational bundle on some parameter space 
and calculate the canonical connection form to determine quantum 
holonomies (this is our main result, see sect. 3.3). 
But we could not calculate the curvature form because of being too  
complicated.
 
Moreover we studies several conditions to obtain the universality of 
our model. A condition seems to be lacking. To overcome this point we 
propose an idea that not only usual holonomy but also higher-dimensional 
holonomies (!) must be introduced in our Holonomic Quantum Computation. 
Further study will be required. 

It is not easy to predict the future of geometric quantum computations.
However it is an arena worth challenging for mathematical physicists.

\section{Mathematical Foundation of Quantum Holonomy}

We start with mathematical preliminaries.
Let $\calh$ be a separable Hilbert space over $\fukuso$.
For $m\in{\bf N}$, we set
\begin{equation}
  \label{eq:stmh}
  \kansu{\stm}{\cal H}
  \equiv
  \left\{
    V=\left( v_1,\cdots,v_m\right)
    \in
    \calh\times\cdots\times\calh\vert V^\dagger V=1_m\right\}\ ,
\end{equation}
where $1_m$ is a unit matrix in $\kansu{M}{m,\fukuso}$.
This is called a (universal) Stiefel manifold.
Note that the unitary group $U(m)$ acts on $\kansu{\stm}{\calh}$
from the right:
\begin{equation}
  \label{eq:stmsha}
  \kansu{\stm}{\calh}\times\kansu{U}{m}
  \rightarrow
  \kansu{\stm}{\calh}: \left( V,a\right)\mapsto Va.
\end{equation}
Next we define a (universal) Grassmann manifold
\begin{equation}
  \kansu{\grm}{\calh}
  \equiv
  \left\{
    X\in\kansu{M}{\calh}\vert
    X^2=X, X^\dagger=X\  \mathrm{and}\  \mathrm{tr}X=m\right\}\ ,
\end{equation}
where $M(\calh)$ denotes a space of all bounded linear operators on $\calh$.
Then we have a projection
\begin{equation}
  \label{eq:piteigi}
  \pi : \kansu{\stm}{\calh}\rightarrow\kansu{\grm}{\calh}\ ,
  \quad \kansu{\pi}{V}\equiv VV^\dagger\ ,
\end{equation}
compatible with the action (\ref{eq:stmsha}) 
($\kansu{\pi}{Va}=Va(Va)^\dagger=Vaa^\dagger V^\dagger=VV^\dagger=
\kansu{\pi}{V}$).

Now the set
\begin{equation}
  \label{eq:principal}
  \left\{
    \kansu{U}{m}, \kansu{\stm}{\calh}, \pi, \kansu{\grm}{\calh}
  \right\}\ ,
\end{equation}
is called a (universal) principal $U(m)$ bundle, 
see \cite{MN} and \cite{KF1}.\quad We set
\begin{equation}
  \label{eq:emh}
  \kansu{\eem}{\cal H}
  \equiv
  \left\{
    \left(X,v\right)
    \in
    \kansu{\grm}{\calh}\times\calh \vert Xv=v \right\}\ .
\end{equation}
Then we have also a projection 
\begin{equation}
  \label{eq:piemgrm}
  \pi : \kansu{\eem}{\calh}\rightarrow\kansu{\grm}{\calh}\ ,
  \quad \kansu{\pi}{\left(X,v\right)}\equiv X\ .
\end{equation}
The set
\begin{equation}
  \label{eq:universal}
  \left\{
    \fukuso^m, \kansu{\eem}{\calh}, \pi, \kansu{\grm}{\calh}
  \right\}\ ,
\end{equation}
is called a (universal) $m$-th vector bundle. This vector bundle is 
one associated with the principal $U(m)$ bundle (\ref{eq:principal})
.

Next let $M$ be a finite or infinite dimensional differentiable manifold 
and the map $P : M \rightarrow \kansu{\grm}{\calh}$ be given (called a 
projector). Using this $P$ we can make 
the bundles (\ref{eq:principal}) and (\ref{eq:universal}) pullback 
over $M$ :
\begin{eqnarray}
  \label{eq:hikimodoshi1}
  &&\left\{\kansu{U}{m},\widetilde{St}, \pi_{\widetilde{St}}, M\right\}
  \equiv
  P^*\left\{\kansu{U}{m}, \kansu{\stm}{\calh}, \pi, 
  \kansu{\grm}{\calh}\right\}
  \ , \\
  \label{eq:hikimodoshi2}
  &&\left\{\fukuso^m,\widetilde{E}, \pi_{\widetilde{E}}, M\right\}
  \equiv
  P^*\left\{\fukuso^m, \kansu{\eem}{\calh}, \pi, \kansu{\grm}{\calh}\right\}
  \ ,
\end{eqnarray}
see \cite{MN}. (\ref{eq:hikimodoshi2}) is of course a vector bundle 
associated with (\ref{eq:hikimodoshi1}).

Let $\calm$ be a parameter space and we denote by $\lam$ its element. 
Let $\slam$ be a fixed reference point of $\calm$. Let $H_\lam$ be 
a family of Hamiltonians parameterized by $\calm$ which act on a Fock space 
$\calh$. We set $H_0$ = $H_\slam$ for simplicity and assume that this has 
a $m$-fold degenerate vacuum :
\begin{equation}
  H_{0}v_j = \mathbf{0},\quad j = 1 \sim m. 
\end{equation}
These $v_j$'s form a $m$-dimensional vector space. We may assume that 
$\langle v_{i}\vert v_{j}\rangle = \delta_{ij}$. Then $\left(v_1,\cdots,v_m
\right) \in \kansu{\stm}{\calh}$ and 
\[
  F_0 \equiv \left\{\sum_{j=1}^{m}x_{j}v_{j}\vert x_j \in \fukuso \right\} 
  \cong \fukuso^m.
\]
Namely, $F_0$ is a vector space associated with o.n.basis 
$\left(v_1,\cdots,v_m\right)$.

Next we assume for simplicity 
that a family of unitary operators parameterized by $\calm$
\begin{equation}
  \label{eq:ufamily} 
  W : \calm \rightarrow U(\calh),\quad W(\slam) = {\rm id}.
\end{equation}
is given and $H_{\lam}$ above is given by the following isospectral family
\begin{equation}
 H_{\lam} \equiv W(\lam)H_0 W(\lam)^{-1}.
\end{equation}
In this case there is no level crossing of eigenvalues. Making use of 
$W(\lam)$ we can define a projector
\begin{equation}
  \label{eq:pfamily}
 P : \calm \rightarrow \kansu{\grm}{\calh}, \quad 
 P(\lam) \equiv
  W(\lam) \left(\sum^{m}_{j=1}v_{j}v_{j}^{\dagger}\right)W(\lam)^{-1}
\end{equation}
and have the pullback bundles over $\calm$
\begin{equation}
  \label{eq:target}
 \left\{\kansu{U}{m},\widetilde{St}, \pi_{\widetilde{St}}, \calm\right\},\quad 
 \left\{\fukuso^m,\widetilde{E}, \pi_{\widetilde{E}}, \calm\right\}.
\end{equation}

For the latter we set
\begin{equation}
  \label{eq:vacuum}
 \ket{vac} = \left(v_1,\cdots,v_m\right).
\end{equation}
In this case a canonical connection form $\cala$ of 
$\left\{\kansu{U}{m},\widetilde{St}, \pi_{\widetilde{St}}, \calm\right\}$ is 
given by 
\begin{equation}
  \label{eq:cform}
 \cala = \bra{vac}W(\lam)^{-1}d W(\lam)\ket{vac},
\end{equation}
where $d$ is a differential form on $\calm$, and its curvature form by
\begin{equation}
  \label{eq:curvature}
  \calf \equiv d\cala+\cala\wedge\cala,
\end{equation}
see \cite{SW} and \cite{MN}.

Let $\gamma$ be a loop in $\calm$ at $\slam$., $\gamma : [0,1] 
\rightarrow \calm, \gamma(0) = \gamma(1)$. For this $\gamma$ a holonomy 
operator $\Gamma_{\cala}$ is defined :
\begin{equation}
  \label{eq:holonomy}
  \Gamma_{\cala}(\gamma) = {\cal P}exp\left\{\oint_{\gamma}\cala\right\} 
  \in \kansu{U}{m},
\end{equation}
where ${\cal P}$ means path-ordered. This acts on the fiber $F_0$ at 
$\slam$ of the vector bundle 
$\left\{\fukuso^m,\widetilde{E}, \pi_{\widetilde{E}}, M\right\}$ as follows :
${\textbf x} \rightarrow \Gamma_{\cala}(\gamma){\textbf x}$.\quad 
The holonomy group $Hol(\cala)$ is in general subgroup of $\kansu{U}{m}$ 
. In the case of $Hol(\cala) = \kansu{U}{m}$,   $\cala$ is called 
irreducible, see \cite{MN}.

In the Holonomic Quantum Computer we take  
\begin{eqnarray}
  \label{eq:information}
  &&{\rm Encoding\ of\ Information} \Longrightarrow {\textbf x} \in F_0 , 
  \nonumber \\
  &&{\rm Processing\ of\ Information} \Longrightarrow \Gamma_{\cala}(\gamma) : 
  {\textbf x} \rightarrow \Gamma_{\cala}(\gamma){\textbf x}.
\end{eqnarray}

\section{Holonomic Quantum Computation}

We apply the results of last section to Quantum Optics and discuss  
(optical) Holonomic Quantum Computation proposed by \cite{ZR} and 
\cite{PC}.

\subsection{Holonomic Quantum Computation 1 $\cdots$ \cite{KF2}}

Let $a(a^\dagger)$ be the annihilation (creation) operator of the harmonic 
oscillator.
If we set $N\equiv a^\dagger a$ (:\ number operator), then
\begin{equation}
  [N,a^\dagger]=a^\dagger\ ,\
  [N,a]=-a\ ,\
  [a,a^\dagger]=1\ .
\end{equation}
Let $\calh$ be a Fock space generated by $a$ and $a^\dagger$, and
$\{\ket{n}\vert n\in\futon\cup\{0\}\}$ be its basis.
The actions of $a$ and $a^\dagger$ on $\calh$ are given by
\begin{equation}
  \label{eq:shoukou}
  a\ket{n} = \sqrt{n}\ket{n-1}\ ,\
  a^\dagger\ket{n} = \sqrt{n+1}\ket{n+1}\ ,
\end{equation}
where $\ket{0}$ is a vacuum ($a\ket{0}=0$).

Now we set
\begin{equation}
  \label{eq:kdaisu}
  {\widetilde K}_{+}\equiv{1\over2}\left( a^\dagger\right)^2\ ,\ \ 
  {\widetilde K}_{-}\equiv{1\over2}a^2\ ,\ \ 
  {\widetilde K}_{3}\equiv{1\over2}\left( a^\dagger a +{1\over2}\right)\ ,
\end{equation}
then we have
\begin{equation}
  [{\widetilde K}_{3},{\widetilde K}_{+}]={\widetilde K}_{+}\ ,\
  [{\widetilde K}_{3},{\widetilde K}_{-}]=-{\widetilde K}_{-}\ ,\
  [{\widetilde K}_{+},{\widetilde K}_{-}]=-2{\widetilde K}_{3}\ .
\end{equation}
That is, the set $\{{\widetilde K}_{+},{\widetilde K}_{-},
{\widetilde K}_{3}\}$ gives a unitary representation of $su(1,1)$ with 
spin $1/4$ and $3/4$, \cite{AP}.

In the following we treat unitary coherent operators based on Lie algebras 
$\fukuso$ and $su(1,1)$. 

\noindent{\bfseries Definition}\quad We set 
\begin{eqnarray}
  \label{eq:d-operator}
  {\rm [NC]}&&\ 
D(\alpha) = e^{\alpha a^\dagger-\bar{\alpha}a} \quad 
  {\rm for}\  \alpha \in \fukuso , \\
  \label{eq:s-operator}
  {\rm [NC]}&&\ 
S(\beta) = e^{\beta {\widetilde K}_{+} - \bar{\beta}{\widetilde K}_{-}}\quad 
  {\rm for}\  \beta \in \fukuso.
\end{eqnarray}
For the details of $D(\alpha)$ and $ S(\beta)$ see \cite{AP} and 
\cite{FKSF1}. For the latter convenience let us list well-known 
disentangling formulas.

\noindent{\bfseries Lemma 3-1-1}\quad We have
\begin{eqnarray}
  \label{eq:d-formula}
{\rm [NC]}\ D(\alpha) &=&  
  e^{-\alzeta^2/2}e^{\alpha a^\dagger}e^{-\bar\alpha a},  \\
  \label{eq:s-formula}
{\rm [NC]}\ S(\beta) &=& 
  e^{\zeta {\widetilde K}_{+}}
  e^{\kansu{\log}{1-\vert\zeta\vert^2}{\widetilde K}_{3}}
  e^{-\bar\zeta {\widetilde K}_{-}}\quad \mbox{where}\quad 
          \zeta=\frac{\beta\tanh\bezeta}{\bezeta}
 \end{eqnarray}
As for a generalization of these formulas see \cite{FS}.

Under preliminaries above let us proceed to the main subject. 
Let $H_0$ be a Hamiltonian
\begin{equation}
  \label{eq:ham}
  H_0\equiv\hbar X N(N-1).
\end{equation}

This has a $2$-fold degenerate vacuum because if we set
\begin{equation}
  F_0
  \equiv
  \mathrm{Vect}\left\{\ket{0},\ket{1}\right\}\ ,
\end{equation}
then $H_0{F_0} = 0$. 

Now note that $\ket{vac}\equiv (\ket{0},\ket{1}) \in \kansu{\st}{\calh}$
in (\ref{eq:stmh}). 
We consider a two-parameter isospectral family
\begin{eqnarray}
  \label{eq:hlambda-mu}
  &&H_{(\alpha,\beta)}
  \equiv
  \kansu{O}{\alpha,\beta}H_0\kansu{O}{\alpha,\beta}^{-1}\ ,
  \\
  &&\kansu{O}{\alpha,\beta}
  \equiv
  D(\alpha)S(\beta), \quad O(0,0)= \mbox{id}\ ,
   \label{eq:huni}
\end{eqnarray}
where $(\alpha,\beta) \in \fukuso^2$.
Since (\ref{eq:hlambda-mu}) is isospectral we have no level-crossing of
eigenvalues for the parameters (adiabatic!).
In the following we focus our attention on the $2$-fold degenerate
vacuum.

For this system let us calculate a connection form (\ref{eq:cform})
in the last section. For that we set 
\begin{equation}
  \label{eq:connections}
  A_{\alpha} = \bra{vac}O(\alpha,\beta)^{-1}\frac{\partial}{\partial \alpha}
            O(\alpha,\beta)\ket{vac}, \quad 
  A_{\beta} = \bra{vac}O(\alpha,\beta)^{-1}\frac{\partial}{\partial \beta}
            O(\alpha,\beta)\ket{vac}.
\end{equation}
Here remaking
\begin{eqnarray}
 O(\alpha,\beta)^{-1}\frac{\partial}{\partial \alpha}O(\alpha,\beta) &=& 
 S(\beta)^{-1}\left\{D(\alpha)^{-1}\frac{\partial}{\partial \alpha}
 D(\alpha)\right\}S(\beta), \nonumber \\
 O(\alpha,\beta)^{-1}\frac{\partial}{\partial \beta}O(\alpha,\beta) &=& 
 S(\beta)^{-1}\frac{\partial}{\partial \beta}S(\beta) \nonumber
\end{eqnarray}
and using Lemma 1,

\noindent{\bfseries Lemma 3-1-2}\quad  We have
\begin{eqnarray}
  \label{eq:lemma-2-ue}
  O^{-1}\frac{\partial}{\partial \alpha} O
  &=&
  {\bar\alpha\over2}1+\cosh\bezeta a^\dagger
  +{\bar\beta\sinh\bezeta\over\bezeta}a\ ,
  \\
  O^{-1}\frac{\partial}{\partial \beta} O
  &=&
  {1\over2}
  \left(1+{\sinh(2\bezeta)\over{2\bezeta}}\right)
  {1\over2}\left( a^\dagger\right)^2
  + 
  {{\bar\beta(-1+\cosh(2\bezeta))}\over{2\bezeta^2}}
  {1\over2}\left( a^\dagger a+{1\over2}\right) 
  \nonumber\\
  &&+
  {\bar\beta^2\over{2\bezeta^2}}
  \left(-1+{\sinh(2\bezeta)\over{2\bezeta}}\right)
  {1\over2}a^2\ .
  \label{eq:lemma-2-shita}
\end{eqnarray}
Compare (\ref{eq:lemma-2-ue}) and (\ref{eq:lemma-2-shita})
with those of \cite{ZR}.

From this lemma it is easy to calculate $A_{\lambda}$ and $A_{\mu}$.
Before stating the result let us prepare some notations.

\begin{equation}
  E=\pmatrix{&1\cr0&\cr}\ ,\
  F=\pmatrix{&0\cr1&\cr}\ ,\
  K=\pmatrix{0&\cr&1\cr}\ ,\
  L=\pmatrix{1&\cr&1\cr}\ .
\end{equation}

\noindent{\bfseries Proposition 3-1-3}\quad We have
\begin{eqnarray}
  \label{eq:a-lambda}
  A_\alpha
  &=&
  {\bar\alpha\over2}L+\cosh\bezeta F
  +{\bar\beta\sinh\bezeta\over\bezeta}E\ ,
  \\
  A_\beta
  &=&
  {{\bar\beta(-1+\cosh(2\bezeta))}\over{4\bezeta^2}}
  \left( K+{1\over2}L\right)\ . 
 \label{eq:a-mu}
\end{eqnarray}

A comment here is in order.
By the diagonal parts of (\ref{eq:a-lambda}) and (\ref{eq:a-mu})
we have the Berry phase stated in \cite{SLB} easily.

Since the connection $\cala$ is anti-hermitian ($\cala^\dagger=-\cala$), 
it can be written as
\begin{equation}
  \label{eq:connection-form}
  \cala
  =
  A_\alpha d\alpha + A_\beta d\beta - {A_\alpha}^\dagger d\bar\alpha
  - {A_\beta}^\dagger d\bar\beta\ ,
\end{equation}
so that it's curvature form $\calf = d\cala + \cala\wedge\cala$ becomes
\begin{eqnarray}
  \label{eq:curvature-form}
  \calf
  &=&
  \left(
    \partial_\alpha A_\beta-\partial_\beta A_\alpha
    +[A_\alpha,A_\beta]
  \right) d\alpha\wedge d\beta
  \nonumber\\
  &&
  -\left(
    \partial_\alpha {A_\alpha}^\dagger+\partial_{\bar\alpha}A_\alpha
    +[A_\alpha,{A_\alpha}^\dagger]
  \right) d\alpha\wedge d\bar\alpha
  \nonumber\\
  &&
   -\left(
    \partial_\alpha {A_\beta}^\dagger+\partial_{\bar\beta}A_\alpha
    +[A_\alpha,{A_\beta}^\dagger]
  \right) d\alpha\wedge d\bar\beta
  \nonumber\\
  &&-\left(
    \partial_\beta {A_\alpha}^\dagger+\partial_{\bar\alpha}A_\beta
    +[A_\beta,{A_\alpha}^\dagger]
  \right) d\beta\wedge d\bar\alpha
  \nonumber\\
  &&-\left(
    \partial_\beta {A_\beta}^\dagger+\partial_{\bar\beta}A_\beta
    +[A_\beta,{A_\beta}^\dagger]
  \right) d\beta\wedge d\bar\beta
  \nonumber\\
  &&-\left(
    \partial_{\bar\alpha}{A_\beta}^\dagger-
   \partial_{\bar\beta}{A_\alpha}^\dagger
    +[{A_\beta}^\dagger,{A_\alpha}^\dagger]
  \right) d\bar\alpha\wedge d\bar\beta\ .
\end{eqnarray}

Now let us state our result.

\noindent{\bfseries Theorem 3-1-4}
\begin{eqnarray}
  \calf
  &=&
  \Bigg\{
    {{\bar\beta}^2\cosh\bezeta\over{2\bezeta^2}}
    \left(
      -1+{\sinh(2\bezeta)\over{2\bezeta}}
    \right) E
    -{\bar\beta\sinh\bezeta\over2\bezeta}
    \left(
      1+{\sinh(2\bezeta)\over{2\bezeta}}
    \right) F
  \Bigg\}
  d\alpha\wedge d\beta
  \nonumber\\
  &&
  -2K d\alpha\wedge d\bar\alpha
  \nonumber\\
  &&
  -\Bigg\{
  {\cosh\bezeta\over2}
  \left(1+{\sinh(2\bezeta)\over{2\bezeta}}\right) E
  -{\beta\sinh\bezeta\over2\bezeta}
  \left(-1+{\sinh(2\bezeta)\over{2\bezeta}}\right) F
  \Bigg\} d\alpha\wedge d\bar\beta
  \nonumber\\
  &&
  -\Bigg\{
  -{\bar\beta\sinh\bezeta\over2\bezeta}
  \left(-1+{\sinh(2\bezeta)\over{2\bezeta}}\right) E
  +{\cosh\bezeta\over2}
  \left(1+{\sinh(2\bezeta)\over{2\bezeta}}\right) F
  \Bigg\} d\beta\wedge d\bar\alpha
  \nonumber\\
  &&
  -{\sinh(2\bezeta)\over{2\bezeta}} 
    \left(K+{1\over2}L \right)
  d\beta\wedge d\bar\beta
  \nonumber\\
  &&
  -\Bigg\{
  -{\beta\sinh\bezeta\over2\bezeta}
  \left(1+{\sinh(2\bezeta)\over{2\bezeta}}\right) E
  +{\beta^2\cosh\bezeta\over2\bezeta^2}
  \left(-1+{\sinh(2\bezeta)\over{2\bezeta}}\right) F
  \Bigg\}
  d\bar\alpha\wedge d\bar\beta\ .
  \nonumber\\
  &&{}
\end{eqnarray}

It is easy to see that the target of $\calf$ covers all of Lie 
algebra $u(2)$.
This means that the connection $\cala$ is irreducible $\cdots$
the holonomy group of $\cala$ is just $U(2)$. 
See \cite{ZR}, \cite{PZR} and \cite{MN}.

\noindent{\bfseries Corollary 3-1-5}\quad 
$\cala$ is irreducible (\cite{ZR}),

\subsection{Holonomic Quantum Computation 2 $\cdots$ \cite{KF3}}

Next we consider the system of two-harmonic oscillators. If we set
\begin{equation}
  \label{eq:twosystem}
  a_1 = a \otimes 1,\  a_1^{\dagger} = a^{\dagger} \otimes 1;\ 
  a_2 = 1 \otimes a,\  a_2^{\dagger} = 1 \otimes a^{\dagger},
\end{equation}
then it is easy to see 
\begin{equation}
  \label{eq:relations}
 [a_i, a_j] = [a_i^{\dagger}, a_j^{\dagger}] = 0,\ 
 [a_i, a_j^{\dagger}] = \delta_{ij}, \quad i, j = 1, 2. 
\end{equation}
We also denote by $N_{i} = a_i^{\dagger}a_i$ number operators.

Now since we want to consider coherent states based on Lie algebras $su(2)$ 
and $su(1,1)$, we make use of Schwinger's boson method, see \cite{FKSF1}, 
\cite{FKSF2}. Namely if we set 
\begin{eqnarray}
  \label{eq:Jdaisu}
  {\rm [C]}&&\ su(2) :\quad
     J_+ = a_1^{\dagger}a_2,\ J_- = a_2^{\dagger}a_1,\ 
     J_3 = {1\over2}\left(a_1^{\dagger}a_1 - a_2^{\dagger}a_2\right), \\
  \label{eq:Kdaisu}
  {\rm [NC]}&&\ su(1,1) :\quad
     K_+ = a_1^{\dagger}a_2^{\dagger},\ K_- = a_2 a_1,\ 
     K_3 = {1\over2}\left(a_1^{\dagger}a_1 + a_2^{\dagger}a_2  + 1\right),
\end{eqnarray}
then we have
\begin{eqnarray}
  \label{eq:j-relation}
  {\rm [C]}&&\ su(2) :\quad
     [J_3, J_+] = J_+,\ [J_3, J_-] = - J_-,\ [J_+, J_-] = 2J_3, \\
  \label{eq:k-relation}
  {\rm [NC]}&&\ su(1,1) :\quad
     [K_3, K_+] = K_+,\ [K_3, K_-] = - K_-,\ [K_+, K_-] = -2K_3.
\end{eqnarray}

In the following we treat unitary coherent operators based on Lie algebras 
$su(2)$ and $su(1,1)$. 

\noindent{\bfseries Definition}\quad We set 
\begin{eqnarray}
  \label{eq:j-operator}
  {\rm [C]}&&\ 
U(\xi) = e^{\xi a_1^{\dagger}a_2 - \bar{\xi}a_2^{\dagger}a_1}\quad 
  {\rm for}\  \xi \in \fukuso , \\
  \label{eq:k-operator}
  {\rm [NC]}&&\ 
V(\zeta) = e^{\zeta a_1^{\dagger}a_2^{\dagger} - \bar{\zeta}a_2 a_1}\quad 
  {\rm for}\  \zeta \in \fukuso.
\end{eqnarray}
For the details of $U(\xi)$ and $ V(\zeta)$ see \cite{AP} and 
\cite{FKSF1}. For the latter convenience let us list well-known 
disentangling formulas.

\noindent{\bfseries Lemma 3-2-1}\quad We have
\begin{eqnarray}
  \label{eq:j-formula}
{\rm [C]}\ U(\xi) &=& e^{\eta a_1^{\dagger}a_2}
 e^{{\rm log}\left(1 + {\vert \eta \vert}^{2}\right)
    {1\over2}\left(a_1^{\dagger}a_1 - a_2^{\dagger}a_2\right)}
 e^{- \bar{\eta}a_2^{\dagger}a_1}, \quad 
   where\quad  \eta = \frac{\xi {\rm tan}{\vert \xi \vert}}
                           {{\vert \xi \vert}},  \\
  \label{eq:k-formula}
{\rm [NC]}\ V(\zeta) &=& e^{\kappa a_1^{\dagger}a_2^{\dagger}}
 e^{{\rm log}\left(1 - {\vert \kappa \vert}^{2}\right)
    {1\over2}\left(a_1^{\dagger}a_1 + a_2^{\dagger}a_2 + 1\right)}
 e^{- \bar{\kappa}a_2 a_1},\quad 
   where\quad  \kappa = \frac{\zeta {\rm tanh}{\vert \zeta \vert}}
                             {{\vert \zeta \vert}}. 
 \end{eqnarray}
As for a generalization of these formulas see \cite{FS}.

Let $H_0$ be a Hamiltonian with nonlinear interaction produced by 
a Kerr medium., that is  $H_0 = \hbar {\rm X} N(N-1)$, where X is a 
certain constant, see \cite{PC}. The eigenvectors of $H_0$ corresponding 
to $0$ is $\left\{\ket{0},\ket{1}\right\}$, so its eigenspace is 
${\rm Vect}\left\{\ket{0},\ket{1}\right\} \cong \fukuso^2$. We correspond to 
$0 \rightarrow \ket{0},\ 1 \rightarrow \ket{1}$ for a generator of 
Boolean algebra $\left\{0, 1\right\}$. The space 
${\rm Vect}\left\{\ket{0},\ket{1}\right\}$ is called 1-qubit (quantum bit) 
space, see \cite{AS} or  \cite{RP}. 
Since we are considering the system of two particles, the Hamiltonian that 
we treat in the following is 
\begin{equation}
  \label{eq:hamiltonian}
  H_0 = \hbar {\rm X} N_{1}(N_{1}-1) + \hbar {\rm X} N_{2}(N_{2}-1).
\end{equation}
The eigenspace of $0$ of this Hamiltonian becomes therefore 
\begin{equation}
  \label{eq:eigenspace}
   F_0 = {\rm Vect}\left\{\ket{0},\ket{1}\right\}\otimes 
         {\rm Vect}\left\{\ket{0},\ket{1}\right\} 
         \cong \fukuso^2\otimes \fukuso^2.
\end{equation}
We denote the basis of $F_0$ as $\left\{\kett{0}{0},\kett{0}{1}, \kett{1}{0},
 \kett{1}{1}\right\}$ and set 
\begin{equation}
  \label{eq:2-vacuum}
 \ket{vac} = \left(\kett{0}{0}, \kett{0}{1}, \kett{1}{0}, \kett{1}{1} \right).
\end{equation}

Next we consider the following isospectral family of $H_0$ above :
\begin{eqnarray}
  \label{eq:twofamily}
   H_{(\xi,\zeta)}&=& W(\xi,\zeta)H_0 W(\xi,\zeta)^{-1},\\
  \label{eq:double}
   W(\xi,\zeta) &=&  U(\xi)V(\zeta) \in U(\calh\otimes \calh),\quad 
   W(0,0) = {\rm id}.
\end{eqnarray}
where $(\xi,\zeta) \in \fukuso^2$. 
For this system let us calculate a connection form (\ref{eq:cform})
in the last section. For that we set 
\begin{equation}
  \label{eq:coefficients}
  A_{\xi} = \bra{vac}W(\xi,\zeta)^{-1}\frac{\partial}{\partial \xi}
            W(\xi,\zeta)\ket{vac}, \quad 
  A_{\zeta} = \bra{vac}W(\xi,\zeta)^{-1}\frac{\partial}{\partial \zeta}
            W(\xi,\zeta)\ket{vac}.
\end{equation}
Here remaking
\begin{eqnarray}
 W(\xi,\zeta)^{-1}\frac{\partial}{\partial \xi}W(\xi,\zeta) &=& 
 V(\xi)^{-1}\left\{U(\xi)^{-1}\frac{\partial}{\partial \xi}U(\xi)\right\}
 V(\xi), \nonumber \\
 W(\xi,\zeta)^{-1}\frac{\partial}{\partial \zeta}W(\xi,\zeta) &=& 
 V(\xi)^{-1}\frac{\partial}{\partial \zeta}V(\xi) \nonumber
\end{eqnarray}
and using Lemma 1,
 
\noindent{\bfseries Lemma 3-2-2}\quad we have
\begin{eqnarray}
 &&W^{-1}\frac{\partial}{\partial \xi}W   \nonumber \\
 &=& 
 {1\over2}\left(1+{\sin(2\xizeta)\over2\xizeta}\right)
 \left\{
 \cosh(2\zezeta)a_1^{\dagger}a_2 + 
 {\zeta\sinh(2\zezeta)\over2\zezeta}\left(a_1^{\dagger}\right)^2 +
 {{\bar \zeta}\sinh(2\zezeta)\over2\zezeta}\left(a_2\right)^2 
 \right\} \nonumber \\
 &+& {{\bar \xi}\over2\xizeta^{2}}\left(1-\cos(2\xizeta)\right)
  {1\over2}\left(a_1^{\dagger}a_1 -  a_2^{\dagger}a_2\right) \nonumber \\
 &+&{{\bar \xi}^{2}\over2\xizeta^{2}}
 \left(-1+{\sin(2\xizeta)\over2\xizeta}\right)
 \left\{
 \cosh(2\zezeta)a_2^{\dagger}a_1 + 
 {{\bar \zeta}\sinh(2\zezeta)\over2\zezeta}\left(a_1\right)^2 +
 {\zeta\sinh(2\zezeta)\over2\zezeta}\left(a_2^{\dagger}\right)^2 
 \right\}\ , \nonumber \\
 &&{}  \\
 &&W^{-1}\frac{\partial}{\partial \zeta}W   \nonumber \\
 &=& 
 {1\over2}\left(1+{\sinh(2\zezeta)\over2\zezeta}\right)
  a_1^{\dagger}a_2^{\dagger} 
  + {{\bar \zeta}\over2\zezeta^{2}}\left(-1+ \cosh(2\zezeta)\right)
  {1\over2}\left(a_1^{\dagger}a_1 + a_2^{\dagger}a_2 + 1\right) \nonumber \\
 &+&{{\bar \zeta}^{2}\over2\zezeta^{2}}
 \left(-1+{\sinh(2\zezeta)\over2\zezeta}\right)a_1a_2.
\end{eqnarray}
From this lemma it is easy to calculate $A_{\xi}$ and $A_{\zeta}$.
Before stating the result let us prepare some notations.
\begin{equation}
{\widehat E} = 
\left(
  \begin{array}{cccc}
    0& 0& 0& 0 \\
    0& 0& 1& 0 \\
    0& 0& 0& 0 \\
    0& 0& 0& 0 
  \end{array}
\right), 
{\widehat F} = 
\left(
  \begin{array}{cccc}
    0& 0& 0& 0 \\
    0& 0& 0& 0 \\
    0& 1& 0& 0 \\
    0& 0& 0& 0 
  \end{array}
\right), 
{\widehat H} = 
\left(
  \begin{array}{cccc}
    0& 0& 0& 0 \\
    0& {1\over2}& 0& 0 \\
    0& 0& -{1\over2}& 0 \\
    0& 0& 0& 0 
  \end{array}
\right).
\end{equation}
\begin{equation}
{\widehat A} = 
\left(
  \begin{array}{cccc}
    0& 0& 0& 1 \\
    0& 0& 0& 0 \\
    0& 0& 0& 0 \\
    0& 0& 0& 0 
  \end{array}
\right), 
{\widehat C} = 
\left(
  \begin{array}{cccc}
    0& 0& 0& 0 \\
    0& 0& 0& 0 \\
    0& 0& 0& 0 \\
    1& 0& 0& 0 
  \end{array}
\right), 
{\widehat B} = 
\left(
  \begin{array}{cccc}
    {1\over2}& 0& 0& 0 \\
    0& 1& 0& 0 \\
    0& 0& 1& 0 \\
    0& 0& 0& {3\over2}
  \end{array}
\right).
\end{equation}

\noindent{\bfseries Proposition 3-2-3}\quad We have
\begin{eqnarray}
 A_{\xi}&=& 
 {1\over2}\left(1+{\sin(2\xizeta)\over2\xizeta}\right)\cosh(2\zezeta)
 {\widehat F} 
  - {{\bar \xi}\over2\xizeta^{2}}\left(1-\cos(2\xizeta)\right){\widehat H} 
 \nonumber \\
  &&+ {{\bar \xi}^{2}\over2\xizeta^{2}}
 \left(-1+{\sin(2\xizeta)\over2\xizeta}\right)\cosh(2\zezeta){\widehat E}, \\
A_{\zeta} &=&  
 {1\over2}\left(1+{\sinh(2\zezeta)\over2\zezeta}\right){\widehat C}
  + 
 {{\bar \zeta}\over2\zezeta^{2}}\left(-1 + \cosh(2\zezeta)\right){\widehat B}
 \nonumber \\ 
  &&+ {{\bar \zeta}^{2}\over2\zezeta^{2}}
 \left(-1+{\sinh(2\zezeta)\over2\zezeta}\right){\widehat A}.
 \end{eqnarray}
Since the connection form $\cala$ is anti-hermitian 
($\cala^{\dagger}=-\cala$), 
it can be written as
\begin{equation}
 \label{eq:calaa}
  \cala =
  A_\xi d\xi + A_\zeta d\zeta - {A_\xi}^{\dagger} d{\bar \xi}
  -{A_\zeta}^{\dagger} d{\bar \zeta}\ ,
\end{equation}
so that it's curvature form $\calf = d\cala + \cala\wedge\cala$ becomes
\begin{eqnarray}
  \label{eq:explicit}
  \calf
  &=&
  \left(
    \partial_\xi A_\zeta-\partial_\zeta A_\xi
    +[A_\xi,A_\zeta]
  \right) d\xi\wedge d\zeta
  \nonumber\\
  &&
  -\left(
    \partial_\xi {A_\xi}^\dagger+\partial_{\bar\xi}A_\xi
    +[A_\xi,{A_\xi}^\dagger]
  \right) d\xi\wedge d\bar\xi
  \nonumber\\
  &&
   -\left(
    \partial_\xi {A_\zeta}^\dagger+\partial_{\bar\zeta}A_\xi
    +[A_\xi,{A_\zeta}^\dagger]
  \right) d\xi\wedge d\bar\zeta
  \nonumber\\
  &&-\left(
    \partial_\zeta {A_\xi}^\dagger+\partial_{\bar\xi}A_\zeta
    +[A_\zeta,{A_\xi}^\dagger]
  \right) d\zeta\wedge d\bar\xi
  \nonumber\\
  &&-\left(
    \partial_\zeta {A_\zeta}^\dagger+\partial_{\bar\zeta}A_\zeta
    +[A_\zeta,{A_\zeta}^\dagger]
  \right) d\zeta\wedge d\bar\zeta
  \nonumber\\
  &&-\left(
    \partial_{\bar\xi}{A_\zeta}^\dagger-\partial_{\bar\zeta}{A_\xi}^\dagger
    +[{A_\zeta}^\dagger,{A_\xi}^\dagger]
  \right) d\bar\xi\wedge d\bar\zeta .
\end{eqnarray}

Now we state our main result.

\noindent{\bfseries Theorem 3-2-4}
\begin{eqnarray}
  \label{eq:main-result}
&& \calf =                \nonumber\\
  &&
  -\Bigg\{
 \left(1+{\sin(2\xizeta)\over2\xizeta}\right)
 {{\bar \zeta}\sinh(2\zezeta)\over2\zezeta}  
 {\widehat F} + 
 {{\bar \xi}^{2}\over\xizeta^{2}}
 \left(-1+{\sin(2\xizeta)\over2\xizeta}\right)
 {{\bar \zeta}\sinh(2\zezeta)\over2\zezeta}  
 {\widehat E}
     \Bigg\}
  d\xi\wedge d\zeta
  \nonumber\\
  &&
  -\Bigg\{
  {\xi\over\xizeta^{2}}
  \left(-1+\cos(2\xizeta)\right)\cosh(2\zezeta){\widehat F}
  -{\sin(2\xizeta)\over\xizeta}
   \left(1+\cosh^2(2\zezeta)\right){\widehat H} \nonumber\\
 &&\ \ \ \ \ +{{\bar\xi}\over\xizeta^{2}}
 \left(-1+\cos(2\xizeta)\right)\cosh(2\zezeta){\widehat E}
     \Bigg\}
  d\xi\wedge d\bar\xi
  \nonumber\\
  &&
  -\Bigg\{
 \left(1+{\sin(2\xizeta)\over2\xizeta}\right)
 {\zeta\sinh(2\zezeta)\over2\zezeta}{\widehat F}  
 + 
 {{\bar\xi}^{2}\over\xizeta^{2}}
 \left(-1+{\sin(2\xizeta)\over2\xizeta}\right)
 {\zeta\sinh(2\zezeta)\over2\zezeta}{\widehat E}
     \Bigg\}
  d\xi\wedge d\bar\zeta
 \nonumber\\
  &&
  -\Bigg\{
 \left(1+{\sin(2\xizeta)\over2\xizeta}\right)
 {{\bar\zeta}\sinh(2\zezeta)\over2\zezeta}{\widehat E}  
 +
 {\xi^{2}\over\xizeta^{2}}
 \left(-1+{\sin(2\xizeta)\over2\xizeta}\right)
 {{\bar\zeta}\sinh(2\zezeta)\over2\zezeta}{\widehat F}
     \Bigg\}
  d\zeta\wedge d\bar\xi
 \nonumber\\
  &&
 - {\sinh(2\zezeta)\over\zezeta}\left(2{\widehat B}-{\textbf 1}_4\right)
  d\zeta\wedge d\bar\zeta
  \nonumber\\
  &&
  +\Bigg\{
 \left(1+{\sin(2\xizeta)\over2\xizeta}\right)
 {\zeta\sinh(2\zezeta)\over2\zezeta}{\widehat E}  
 +
 {\xi^{2}\over\xizeta^{2}}
 \left(-1+{\sin(2\xizeta)\over2\xizeta}\right)
 {\zeta\sinh(2\zezeta)\over2\zezeta}{\widehat F}
     \Bigg\}
  d\bar\xi\wedge d\bar\zeta.
  \nonumber\\
  &&
\end{eqnarray}
From this and the theorem of Ambrose--Singer (see \cite{MN}) 
it is easy to see that 

\noindent{\bfseries Corollary 3-2-5}
\begin{equation}
  Hol(\cala) = SU(2)\times U(1)\ \subset \ U(4).
\end{equation}
Therefore $\cala$ is not irreducible.

\subsection{Holonomic Quantum Computation 3 $\cdots$ Main Result}

The Hamiltonian that we treat in this section is (\ref{eq:hamiltonian})., 
namely 
\[
  H_0 = \hbar {\rm X} N_{1}(N_{1}-1) + \hbar {\rm X} N_{2}(N_{2}-1)
\]
and we consider the following (full) isospectral family :
\begin{eqnarray}
  \label{eq:sixfamily}
   H_{(\alpha_1,\beta_1,\xi,\zeta,\alpha_2,\beta_2)}&=& 
   Z(\alpha_1,\beta_1,\xi,\zeta,\alpha_2,\beta_2)H_0 
   Z(\alpha_1,\beta_1,\xi,\zeta,\alpha_2,\beta_2)^{-1},\\
  \label{eq:triple}
   Z(\alpha_1,\beta_1,\xi,\zeta,\alpha_2,\beta_2) 
   &=& O_{1}(\alpha_1,\beta_1)W(\xi,\zeta)O_{2}(\alpha_2,\beta_2)
    \in U(\calh\otimes \calh), \nonumber \\
   &{}&  Z(0,0,0,0,0,0) = {\rm id}, 
\end{eqnarray}
where $(\alpha_1,\beta_1,\xi,\zeta,\alpha_2,\beta_2) \in \fukuso^6$ and 
$O(\alpha,\beta)$ and $W(\xi,\zeta)$ are respectively (\ref{eq:huni}) 
and (\ref{eq:double}). Namely we consider a family of six-parameters. 
For this system we want to calculate a connection form (\ref{eq:cform}).
For that we set  for simplicity 
\begin{equation}
  \label{eq:manycoefficients}
  A_{\chi} = \bra{vac}Z(\alpha_1,\beta_1,\xi,\zeta,\alpha_2,\beta_2)^{-1}
             \frac{\partial}{\partial \chi}
             Z(\alpha_1,\beta_1,\xi,\zeta,\alpha_2,\beta_2)\ket{vac},
\end{equation}
where $\chi = \alpha_1,\beta_1,\xi,\zeta,\alpha_2,\beta_2$ respectively and   
$\ket{vac}$ is just (\ref{eq:2-vacuum}).

We note 
\begin{eqnarray}
  \label{eq:cartan-1}
Z^{-1}\frac{\partial}{\partial \chi}Z &=& 
 O_{2}^{-1} W^{-1}\left( O_{1}^{-1}\frac{\partial}{\partial \chi}O_{1}
 \right)WO_{2}\quad 
 \mbox{for}\quad  \chi = \alpha_1,\beta_1\ ,  \\
  \label{eq:cartan-2}
Z^{-1}\frac{\partial}{\partial \chi}Z &=& 
 O_{2}^{-1} \left( W^{-1}\frac{\partial}{\partial \chi}W\right)O_{2}\quad 
 \mbox{for}\quad \chi = \xi,\zeta\ ,   \\
  \label{eq:cartan-3}
Z^{-1}\frac{\partial}{\partial \chi}Z &=& 
 O_{2}^{-1}\frac{\partial}{\partial \chi}O_{2}\quad 
 \mbox{for}\quad \chi = \alpha_2,\beta_2\ ,  
\end{eqnarray}
but we have already calculated the main parts 
$X^{-1}\frac{\partial}{\partial \chi}X$ in sect. 3.1 and sect. 3.2.

First let us determine (\ref{eq:cartan-1}). From Lemma 3-1-2 we have 
\begin{eqnarray}
  Z^{-1}\frac{\partial}{\partial \alpha_{1}} Z
  &=&
  {\bar\alpha_1\over2}1+\cosh\beazeta {O_2}^{-1}W^{-1}{a_1}^\dagger WO_2
  +{\bar\beta_1\sinh\beazeta\over\beazeta}{O_2}^{-1}W^{-1}{a_1}WO_2\ ,
  \\
  Z^{-1}\frac{\partial}{\partial \beta_1} Z
  &=&
  {1\over2}
  \left(1+{\sinh(2\beazeta)\over{2\beazeta}}\right)
  {1\over2}\left({O_2}^{-1}W^{-1}{a_1}^\dagger WO_2 \right)^2
  \nonumber \\
  &&+ 
  {{\bar\beta_1(-1+\cosh(2\beazeta))}\over{2\beazeta^2}}
  {1\over2}\Bigg\{ \left({O_2}^{-1}W^{-1}{a_1}^\dagger WO_2 \right)
   \left({O_2}^{-1}W^{-1}{a_1}WO_2 \right)+{1\over2} \Bigg\} 
  \nonumber\\
  &&+
  {\bar\beta_{1}^2\over{2\beazeta^2}}
  \left(-1+{\sinh(2\beazeta)\over{2\beazeta}}\right)
  {1\over2}\left({O_2}^{-1}W^{-1}{a_1}WO_2 \right)^2\ .
\end{eqnarray}

\par \noindent If we set for simplicity
\begin{equation}
 \label{eq:kankeishiki-1}
 O_{2}^{-1}W^{-1}a_{1}WO_{2} = c_{0}+ c_{1}a_{1}+ c_{3}{a_{1}}^{\dagger} 
       + c_{2}a_{2}+ c_{4}{a_{2}}^{\dagger}, 
\end{equation}
then we have 
\begin{eqnarray}
 \label{eq:keisuu-tachi-1} 
   c_{0}&=& \frac{\xi\sin(\xizeta)}{\xizeta}\cosh(\zezeta)\alpha_2 
     + \cos(\xizeta)\frac{\zeta\sin(\zezeta)}{\zezeta}\bar{\alpha}_2 \ , \\ 
   c_{1}&=& \cos(\xizeta)\cosh(\zezeta) \ , \\ 
   c_{3}&=& \frac{\xi\sin(\xizeta)}{\xizeta}
     \frac{\zeta\sin(\zezeta)}{\zezeta} \ ,    \\ 
   c_{2}&=& \Bigg\{
     \frac{\xi\sin(\xizeta)}{\xizeta}\cosh(\zezeta)\cosh(\vert\beta_{2}\vert) +
     \cos(\xizeta)\frac{\zeta\sinh(\zezeta)}{\zezeta}
     \frac{\bar{\beta}_{2}\sinh(\vert\beta_{2}\vert)}{\vert\beta_{2}\vert}
     \Bigg\} \ , \\ 
   c_{4}&=& \Bigg\{
     \frac{\xi\sin(\xizeta)}{\xizeta}\cosh(\zezeta)
      \frac{\beta_{2}\sinh(\vert\beta_{2}\vert)}{\vert\beta_{2}\vert}+ 
     \cos(\xizeta)\frac{\zeta\sinh(\zezeta)}{\zezeta}\cosh(\vert\beta_{2}\vert)
            \Bigg\}\ . 
\end{eqnarray}
See the appendix A. Making use of these 

\noindent{\bfseries Lemma 3-3-1-(i)}\quad we have
\begin{eqnarray}
  \label{eq:cartan-alpha1}
 &&Z^{-1}\frac{\partial}{\partial \alpha_1}Z =
 \Bigg\{
   \frac{\bar{\alpha}_1}{2}+ \cosh(\vert\beta_{1}\vert)\bar{c}_0+ 
   \frac{\bar{\beta}_{1}\sinh(\vert\beta_{1}\vert)}{\vert\beta_{1}\vert}c_0
 \Bigg\}{\rm 1}  \nonumber  \\ 
 &&+  
 \Bigg\{ 
   \cosh(\vert\beta_{1}\vert)\bar{c}_3+ 
   \frac{\bar{\beta}_{1}\sinh(\vert\beta_{1}\vert)}{\vert\beta_{1}\vert}c_1
 \Bigg\}a_1+ 
 \Bigg\{ 
   \cosh(\vert\beta_{1}\vert)\bar{c}_1+ 
   \frac{\bar{\beta}_{1}\sinh(\vert\beta_{1}\vert)}{\vert\beta_{1}\vert}c_3
 \Bigg\}{a_1}^{\dagger}  \nonumber  \\ 
 &&+
 \Bigg\{ 
   \cosh(\vert\beta_{1}\vert)\bar{c}_4+ 
   \frac{\bar{\beta}_{1}\sinh(\vert\beta_{1}\vert)}{\vert\beta_{1}\vert}c_2
 \Bigg\}a_2+ 
 \Bigg\{ 
   \cosh(\vert\beta_{1}\vert)\bar{c}_2+ 
   \frac{\bar{\beta}_{1}\sinh(\vert\beta_{1}\vert)}{\vert\beta_{1}\vert}c_4
 \Bigg\}{a_2}^{\dagger}\ , \\
 &&{} \nonumber  \\ 
  \label{eq:cartan-beta1}
 &&Z^{-1}\frac{\partial}{\partial \beta_1}Z   \nonumber \\
&=&
 \Bigg\{
   {1\over2}\left(1+ \frac{\sinh(2\beazeta)}{2\beazeta}\right)
   {1\over2}\left({\bar{c}_{0}}^2+ \bar{c}_{1}\bar{c}_{3}+ 
   \bar{c}_{2}\bar{c}_{4}\right)+ 
  \frac{\bar{\beta}_{1}}{2\beazeta^2}\left(-1+\cosh(2\beazeta)\right)
  \nonumber \\
 &&\times  
   {1\over2}\left({\vert c_{0}\vert}^2+ {\vert c_{3}\vert}^2+ 
       {\vert c_{4}\vert}^2+ {1\over2} \right)
 +  
  \frac{{\bar{\beta}_{1}}^2}{2\beazeta^2}
   \left(-1+\frac{\sinh(2\beazeta)}{2\beazeta}\right) 
   {1\over2}\left({c_{0}}^2+ c_{1}c_{3}+ c_{2}c_{4}\right) 
\Bigg\}{\rm 1} \nonumber \\
&&+\ 
 \Bigg\{
   {1\over2}\left(1+ \frac{\sinh(2\beazeta)}{2\beazeta}\right)
   \bar{c}_{0}\bar{c}_{3}+ 
  \frac{\bar{\beta}_{1}}{2\beazeta^2}\left(-1+\cosh(2\beazeta)\right)
   {1\over2}\left(c_{1}\bar{c}_{0}+ c_{0}\bar{c}_{3}\right) 
\nonumber \\
&&\quad\ +  
  \frac{{\bar{\beta}_{1}}^2}{2\beazeta^2}
   \left(-1+\frac{\sinh(2\beazeta)}{2\beazeta}\right) 
   c_{0}c_{1} 
\Bigg\}a_1  \nonumber \\ 
&&+\ 
 \Bigg\{
   {1\over2}\left(1+ \frac{\sinh(2\beazeta)}{2\beazeta}\right)
   \bar{c}_{0}\bar{c}_{1}+ 
  \frac{\bar{\beta}_{1}}{2\beazeta^2}\left(-1+\cosh(2\beazeta)\right)
   {1\over2}\left(c_{3}\bar{c}_{0}+ c_{0}\bar{c}_{1}\right)
\nonumber \\
&&\quad\ +  
  \frac{{\bar{\beta}_{1}}^2}{2\beazeta^2}
   \left(-1+\frac{\sinh(2\beazeta)}{2\beazeta}\right) 
   c_{0}c_{3} 
\Bigg\}{a_1}^{\dagger} \nonumber \\ 
&&+\ 
 \Bigg\{
   {1\over2}\left(1+ \frac{\sinh(2\beazeta)}{2\beazeta}\right)
   \bar{c}_{0}\bar{c}_{4}+ 
  \frac{\bar{\beta}_{1}}{2\beazeta^2}\left(-1+\cosh(2\beazeta)\right)
   {1\over2}\left(c_{2}\bar{c}_{0}+ c_{0}\bar{c}_{4}\right) 
\nonumber \\
&&\quad\ +  
  \frac{{\bar{\beta}_{1}}^2}{2\beazeta^2}
   \left(-1+\frac{\sinh(2\beazeta)}{2\beazeta}\right) 
   c_{0}c_{2} 
\Bigg\}a_2  \nonumber \\ 
&&+\ 
 \Bigg\{
   {1\over2}\left(1+ \frac{\sinh(2\beazeta)}{2\beazeta}\right)
   \bar{c}_{0}\bar{c}_{2}+ 
  \frac{\bar{\beta}_{1}}{2\beazeta^2}\left(-1+\cosh(2\beazeta)\right)
   {1\over2}\left(c_{4}\bar{c}_{0}+ c_{0}\bar{c}_{2}\right)
\nonumber \\
&&\quad\ +  
  \frac{{\bar{\beta}_{1}}^2}{2\beazeta^2}
   \left(-1+\frac{\sinh(2\beazeta)}{2\beazeta}\right) 
   c_{0}c_{4} 
\Bigg\}{a_2}^{\dagger}  \nonumber \\ 
&&+\ 
 \Bigg\{\cdots \Bigg\}{a_1}^{2}+\quad  
 \Bigg\{\cdots \Bigg\}\left({a_1}^{\dagger}\right)^{2}+\quad  
 \Bigg\{\cdots \Bigg\}{a_2}^{2}+\quad  
 \Bigg\{\cdots \Bigg\}\left({a_2}^{\dagger}\right)^{2}
 \nonumber \\ 
&&+\ 
 \Bigg\{
   {1\over2}\left(1+ \frac{\sinh(2\beazeta)}{2\beazeta}\right)
   \bar{c}_{3}\bar{c}_{4}+ 
  \frac{\bar{\beta}_{1}}{2\beazeta^2}\left(-1+\cosh(2\beazeta)\right)
   {1\over2}\left(c_{2}\bar{c}_{3}+ c_{1}\bar{c}_{4}\right)
\nonumber \\
&&\quad\ +  
  \frac{{\bar{\beta}_{1}}^2}{2\beazeta^2}
   \left(-1+\frac{\sinh(2\beazeta)}{2\beazeta}\right) 
   c_{1}c_{2} 
\Bigg\}a_1a_2  \nonumber \\ 
&&+\ 
 \Bigg\{
   {1\over2}\left(1+ \frac{\sinh(2\beazeta)}{2\beazeta}\right)
   \bar{c}_{1}\bar{c}_{2}+ 
  \frac{\bar{\beta}_{1}}{2\beazeta^2}\left(-1+\cosh(2\beazeta)\right)
   {1\over2}\left(c_{4}\bar{c}_{1}+ c_{3}\bar{c}_{2}\right)
\nonumber \\
&&\quad\ +  
  \frac{{\bar{\beta}_{1}}^2}{2\beazeta^2}
   \left(-1+\frac{\sinh(2\beazeta)}{2\beazeta}\right) 
   c_{3}c_{4} 
\Bigg\}{a_1}^{\dagger}{a_2}^{\dagger}  \nonumber \\ 
&&+\ 
 \Bigg\{
   {1\over2}\left(1+ \frac{\sinh(2\beazeta)}{2\beazeta}\right)
   \bar{c}_{2}\bar{c}_{3}+ 
  \frac{\bar{\beta}_{1}}{2\beazeta^2}\left(-1+\cosh(2\beazeta)\right)
   {1\over2}\left(c_{4}\bar{c}_{3}+ c_{1}\bar{c}_{2}\right)
\nonumber \\
&&\quad\ +  
  \frac{{\bar{\beta}_{1}}^2}{2\beazeta^2}
   \left(-1+\frac{\sinh(2\beazeta)}{2\beazeta}\right) 
   c_{1}c_{4} 
\Bigg\}a_1{a_2}^{\dagger}  \nonumber \\ 
&&+\ 
 \Bigg\{
   {1\over2}\left(1+ \frac{\sinh(2\beazeta)}{2\beazeta}\right)
   \bar{c}_{1}\bar{c}_{4}+ 
  \frac{\bar{\beta}_{1}}{2\beazeta^2}\left(-1+\cosh(2\beazeta)\right)
   {1\over2}\left(c_{2}\bar{c}_{1}+ c_{3}\bar{c}_{4}\right)
\nonumber \\
&&\quad\ +  
  \frac{{\bar{\beta}_{1}}^2}{2\beazeta^2}
   \left(-1+\frac{\sinh(2\beazeta)}{2\beazeta}\right) 
   c_{2}c_{3} 
\Bigg\}{a_1}^{\dagger}{a_2}  \nonumber \\ 
&&+\ 
 \Bigg\{
   {1\over2}\left(1+ \frac{\sinh(2\beazeta)}{2\beazeta}\right)
   \bar{c}_{1}\bar{c}_{3}+ 
  \frac{\bar{\beta}_{1}}{2\beazeta^2}\left(-1+\cosh(2\beazeta)\right)
   {1\over2}\left({\vert c_{1}\vert}^2+ {\vert c_{3}\vert}^2\right)
\nonumber \\
&&\quad\ +  
  \frac{{\bar{\beta}_{1}}^2}{2\beazeta^2}
   \left(-1+\frac{\sinh(2\beazeta)}{2\beazeta}\right) 
   c_{1}c_{3} 
\Bigg\}{a_1}^{\dagger}{a_1}  \nonumber \\ 
&&+\ 
 \Bigg\{
   {1\over2}\left(1+ \frac{\sinh(2\beazeta)}{2\beazeta}\right)
   \bar{c}_{2}\bar{c}_{4}+ 
  \frac{\bar{\beta}_{1}}{2\beazeta^2}\left(-1+\cosh(2\beazeta)\right)
   {1\over2}\left({\vert c_{2}\vert}^2+ {\vert c_{4}\vert}^2\right)
\nonumber \\
&&\quad\ +  
  \frac{{\bar{\beta}_{1}}^2}{2\beazeta^2}
   \left(-1+\frac{\sinh(2\beazeta)}{2\beazeta}\right) 
   c_{2}c_{4} 
\Bigg\}{a_2}^{\dagger}{a_2} 
\end{eqnarray}
where we have omitted the coefficients of ${a_1}^{2},\  
\left({a_1}^{\dagger}\right)^2,\  {a_2}^{2},\ 
\left({a_1}^{\dagger}\right)^2$  because these terms are neglected as shown 
in the latter.

Next let us determine (\ref{eq:cartan-2}). From Lemma 3-2-2 we have 
\begin{eqnarray}
 &&Z^{-1}\frac{\partial}{\partial \xi}Z  \nonumber \\
 &=& 
 {1\over2}\left(1+{\sin(2\xizeta)\over2\xizeta}\right)
 \left\{
 \cosh(2\zezeta)a_1^{\dagger}{O_2}^{-1}a_2O_2 + 
 {\zeta\sinh(2\zezeta)\over2\zezeta}\left(a_1^{\dagger}\right)^2 +
 {{\bar \zeta}\sinh(2\zezeta)\over2\zezeta}\left({O_2}^{-1}a_2O_2 \right)^2 
 \right\} \nonumber \\
 &+& {{\bar \xi}\over2\xizeta^{2}}\left(1 - \cos(2\xizeta)\right)
  {1\over2}\left\{ a_1^{\dagger}a_1 -  \left({O_2}^{-1}a_2^{\dagger}O_2 \right)
  \left({O_2}^{-1}a_2O_2 \right) \right\}  \nonumber \\
 &+& {{\bar \xi}^{2}\over2\xizeta^{2}}
 \left(-1+{\sin(2\xizeta)\over2\xizeta}\right)
 \left\{
 \cosh(2\zezeta)\left({O_2}^{-1}a_2^{\dagger}O_2 \right)a_1 + 
 {{\bar \zeta}\sinh(2\zezeta)\over2\zezeta}{a_1}^2 +
 {\zeta\sinh(2\zezeta)\over2\zezeta}\left({O_2}^{-1}a_2^{\dagger}O_2 \right)^2 
 \right\}\ ,   \nonumber \\
 &&{}  \\
 &&Z^{-1}\frac{\partial}{\partial \zeta}Z   \nonumber \\
 &=& 
 {1\over2}\left(1+{\sinh(2\zezeta)\over2\zezeta}\right)
  a_1^{\dagger}\left({O_2}^{-1}a_2^{\dagger}O_2 \right)  \nonumber \\
  &+& {{\bar \zeta}\over2\zezeta^{2}}\left(-1+ \cosh(2\zezeta)\right)
  {1\over2}\left\{a_1^{\dagger}a_1 + 
  \left({O_2}^{-1}a_2^{\dagger}O_2 \right)\left({O_2}^{-1}a_2O_2 \right)
    + 1\right\} \nonumber \\
 &+& {{\bar \zeta}^{2}\over2\zezeta^{2}}
 \left(-1+{\sinh(2\zezeta)\over2\zezeta}\right)
 a_1\left({O_2}^{-1}a_2O_2 \right)\ .
\end{eqnarray}

\par \noindent 
If we set for simplicity
\begin{equation}
 \label{eq:kankeishiki-2}
   O_{2}^{-1}a_{2}O_{2} = \alpha_2 + d_{1}a_{2}+ d_{2}{a_{2}}^{\dagger}, 
\end{equation}
then we have 
\begin{equation}
 \label{eq:keisuu-tachi-2}
      d_{1}= \cosh(\bebzeta),\quad  
      d_{2}= \frac{{\beta_2}\sinh(\bebzeta)}{\bebzeta}\ .
\end{equation}

\noindent Making use of these 

\noindent{\bfseries Lemma 3-3-1-(ii)}\quad we have
\begin{eqnarray}
  \label{eq:cartan-xi}
 &&Z^{-1}\frac{\partial}{\partial \xi}Z   \nonumber \\
&=&
 \Bigg\{
   {1\over2}\left(1+ \frac{\sin(2\xizeta)}{2\xizeta}\right)
   \frac{\bar{\zeta}\sinh(2\zezeta)}{2\zezeta}
   \left({\alpha_{2}}^2+ d_1d_2\right)- 
  \frac{\bar{\xi}}{2\xizeta^2}\left(1-\cos(2\xizeta)\right)
  {1\over2}\left({\vert \alpha_{2}\vert}^2+ {\vert d_{2}\vert}^2 \right)
  \nonumber \\ 
&&+  
  \frac{\bar{\xi}^2}{2\xizeta^2}
   \left(-1+\frac{\sin(2\xizeta)}{2\xizeta}\right)
   \frac{\zeta\sinh(2\zezeta)}{2\zezeta} 
   \left({\bar{\alpha_2}}^2+ \bar{d_1}\bar{d_2}\right) 
\Bigg\}{\rm 1} \nonumber \\
&&+ 
 \Bigg\{
  \frac{\bar{\xi}^2}{2\xizeta^2}
   \left(-1+\frac{\sin(2\xizeta)}{2\xizeta}\right)
   \cosh(2\zezeta)\bar{\alpha_2}
 \Bigg\}a_1 + 
 \Bigg\{
   {1\over2}\left(1+\frac{\sin(2\xizeta)}{2\xizeta}\right)
   \cosh(2\zezeta){\alpha_2}
 \Bigg\}{a_1}^\dagger  \nonumber \\ 
&&+  
 \Bigg\{
   {1\over2}\left(1+ \frac{\sin(2\xizeta)}{2\xizeta}\right)
   \frac{\bar{\zeta}\sinh(2\zezeta)}{2\zezeta}
   2\alpha_2d_1 - 
  \frac{\bar{\xi}}{2\xizeta^2}\left(1-\cos(2\xizeta)\right)
   {1\over2}\left(\bar{\alpha_2}d_1+ \alpha_2\bar{d_2}\right) 
\nonumber \\ 
&&\quad\ +  
  \frac{\bar{\xi}^2}{2\xizeta^2}
   \left(-1+\frac{\sin(2\xizeta)}{2\xizeta}\right)
   \frac{\zeta\sinh(2\zezeta)}{2\zezeta} 
   2\bar{\alpha_2}\bar{d_2} 
\Bigg\}a_2 \nonumber \\
&&+ 
 \Bigg\{
   {1\over2}\left(1+ \frac{\sin(2\xizeta)}{2\xizeta}\right)
   \frac{\bar{\zeta}\sinh(2\zezeta)}{2\zezeta}
   2\alpha_2d_2 - 
  \frac{\bar{\xi}}{2\xizeta^2}\left(1-\cos(2\xizeta)\right)
  {1\over2}\left({\alpha_2}\bar{d_1}+ \bar{\alpha_2}{d_2}\right)
 \nonumber \\  
&&\quad\  +  
  \frac{\bar{\xi}^2}{2\xizeta^2}
   \left(-1+\frac{\sin(2\xizeta)}{2\xizeta}\right)
   \frac{\zeta\sinh(2\zezeta)}{2\zezeta} 
   2\bar{\alpha_2}\bar{d_1} 
\Bigg\}{a_2}^\dagger \nonumber \\
&&+ 
 \Bigg\{
   {1\over2}\left(1+\frac{\sin(2\xizeta)}{2\xizeta}\right)
   \cosh(2\zezeta)d_1
 \Bigg\}{a_1}^\dagger{a_2}+ 
 \Bigg\{
  \frac{\bar{\xi}^2}{2\xizeta^2}
   \left(-1+\frac{\sin(2\xizeta)}{2\xizeta}\right)
   \cosh(2\zezeta)\bar{d_1}
 \Bigg\}a_1{a_2}^\dagger  \nonumber \\ 
&&+ 
 \Bigg\{
  \frac{\bar{\xi}^2}{2\xizeta^2}
   \left(-1+\frac{\sin(2\xizeta)}{2\xizeta}\right)
   \cosh(2\zezeta)\bar{d_2}
 \Bigg\}a_1{a_2}+
 \Bigg\{
   {1\over2}\left(1+\frac{\sin(2\xizeta)}{2\xizeta}\right)
   \cosh(2\zezeta)d_2
 \Bigg\}{a_1}^\dagger{a_2}^\dagger  \nonumber \\ 
&&+ 
 \Bigg\{\cdots \Bigg\}{a_1}^{2}+\quad  
 \Bigg\{\cdots \Bigg\}\left({a_1}^{\dagger}\right)^{2}+\quad  
 \Bigg\{\cdots \Bigg\}{a_2}^{2}+\quad  
 \Bigg\{\cdots \Bigg\}\left({a_2}^{\dagger}\right)^{2}
 \nonumber \\ 
&&+\ 
\frac{\bar{\xi}}{2\xizeta^2}\left(1-\cos(2\xizeta)\right)
{1\over2}{a_1}^\dagger{a_1} \nonumber \\ 
&&+\ 
 \Bigg\{
   {1\over2}\left(1+ \frac{\sin(2\xizeta)}{2\xizeta}\right)
   \frac{\bar{\zeta}\sinh(2\zezeta)}{2\zezeta}
   2d_1d_2- 
  \frac{\bar{\xi}}{2\xizeta^2}\left(1-\cos(2\xizeta)\right) 
   {1\over2}\left({\vert d_{1}\vert}^2+ {\vert d_{2}\vert}^2 \right)
  \nonumber \\ 
&&\quad\ +  
  \frac{\bar{\xi}^2}{2\xizeta^2}
   \left(-1+\frac{\sin(2\xizeta)}{2\xizeta}\right)
   \frac{\zeta\sinh(2\zezeta)}{2\zezeta} 
   2\bar{d_1}\bar{d_2}
\Bigg\}{a_2}^\dagger{a_2}\ , \\
 &&{} \nonumber  \\ 
  \label{eq:cartan-zeta}
 &&Z^{-1}\frac{\partial}{\partial \zeta}Z   \nonumber \\
&=&
 \frac{\bar{\zeta}}{2\zezeta^2}(-1+ \cosh(2\zezeta))
   \left(1+ {\vert \alpha_{2}\vert}^2+ {\vert d_{2}\vert}^2 \right){\rm 1}
\nonumber \\
&&+
 \frac{\bar{\zeta}^2}{2\zezeta^2}   
 \left(-1+\frac{\sinh(2\zezeta)}{2\zezeta}\right){\alpha}_2 a_1 
 +
 {1\over2} \left(1+\frac{\sinh(2\zezeta)}{2\zezeta}\right)
 {\bar{\alpha}}_2 {a_1}^\dagger 
\nonumber \\
&&+
 \frac{\bar{\zeta}}{2\zezeta^2}(-1+ \cosh(2\zezeta))
  \left(\bar{\alpha_2}d_1+ \alpha_2\bar{d_2}\right) a_2 
+
 \frac{\bar{\zeta}}{2\zezeta^2}(-1+ \cosh(2\zezeta))
  \left(\bar{\alpha_2}d_2+ \alpha_2\bar{d_1}\right) {a_2}^\dagger
\nonumber \\
&&+
 \frac{\bar{\zeta}^2}{2\zezeta^2}   
 \left(-1+\frac{\sinh(2\zezeta)}{2\zezeta}\right)d_1 a_1a_2 
 +
 {1\over2} \left(1+\frac{\sinh(2\zezeta)}{2\zezeta}\right)
 {\bar d}_1 {a_1}^\dagger{a_2}^\dagger 
\nonumber \\
&&+
 \frac{\bar{\zeta}^2}{2\zezeta^2}   
 \left(-1+\frac{\sinh(2\zezeta)}{2\zezeta}\right)d_2 a_1{a_2}^\dagger 
 +
 {1\over2} \left(1+\frac{\sinh(2\zezeta)}{2\zezeta}\right)
 {\bar d}_2 {a_1}^\dagger{a_2} 
\nonumber \\
&&+ 
 \Bigg\{\cdots \Bigg\}{a_2}^{2}+\quad  
 \Bigg\{\cdots \Bigg\}\left({a_2}^{\dagger}\right)^{2}
 \nonumber \\ 
&&+ 
 \frac{\bar{\zeta}}{2\zezeta^2}(-1+ \cosh(2\zezeta)){a_1}^\dagger{a_1} 
+
 \frac{\bar{\zeta}}{2\zezeta^2}(-1+ \cosh(2\zezeta))
 \left({\vert d_1\vert}^2+ {\vert d_{2}\vert}^2 \right){a_2}^\dagger{a_2}\ .
\end{eqnarray}

\vspace{5mm}
Last let us determine (\ref{eq:cartan-3}). But we have already calculated 
in Lemma 3-1-2.

\noindent{\bfseries Lemma 3-3-1-(iii)}\quad we have
\begin{eqnarray}
  \label{eq:cartan-alpha2}
 Z^{-1}\frac{\partial}{\partial \alpha_2}Z 
 &=&
  {\bar{\alpha}_2\over2}1+\cosh\bebzeta {a_2}^\dagger
  +{\bar{\beta}_2\sinh\bebzeta\over\bebzeta}{a_2}\ ,
  \\
  \label{eq:cartan-beta2}
 Z^{-1}\frac{\partial}{\partial \beta_2}Z 
 &=&
  {1\over2}
  \left(1+{\sinh(2\bebzeta)\over{2\bebzeta}}\right)
  {1\over2}\left({a_2}^\dagger\right)^2
  + 
  {{\bar{\beta}_2(-1+\cosh(2\bebzeta))}\over{2\bebzeta^2}}
  {1\over2}\left({a_2}^\dagger{a_2}+{1\over2}\right) 
  \nonumber\\
  &&+
  {\bar{\beta}_2^2\over{2\bebzeta^2}}
  \left(-1+{\sinh(2\bebzeta)\over{2\bebzeta}}\right)
  {1\over2}{a_2}^2\ .
\end{eqnarray}

Let us calculate (\ref{eq:manycoefficients}). Since $\ket{vac}=
\left(\kett{0}{0}, \kett{0}{1}, \kett{1}{0}, \kett{1}{1} \right)$ 
we have
\begin{eqnarray}
  && a_1\ket{vac}=\left(0, 0, \kett{0}{0}, \kett{0}{1} \right),\quad 
   {a_1}^\dagger\ket{vac}=\left(\kett{1}{0}, \kett{1}{1},*, *  \right),
        \nonumber \\
  && a_2\ket{vac}=\left(0, \kett{0}{0}, 0, \kett{1}{0} \right),\quad 
   {a_2}^\dagger\ket{vac}=\left(\kett{0}{1},*, \kett{1}{1}, * \right),
        \nonumber \\
  && {a_1}^2\ket{vac}=\left(0, 0, 0, 0 \right),\quad 
   \left({a_1}^\dagger\right)^2\ket{vac}=\left(*, *, *, *  \right),
        \nonumber \\
  && {a_2}^2\ket{vac}=\left(0, 0, 0, 0 \right),\quad 
   \left({a_2}^\dagger\right)^2\ket{vac}=\left(*, *, *, *  \right),
        \nonumber \\
  &&{a_1}{a_2}\ket{vac}=\left(0, 0, 0, \kett{0}{0} \right),\quad 
    {a_1}^\dagger{a_2}^\dagger\ket{vac}=\left(\kett{1}{1}, *, *, * \right)
        \nonumber \\
  && {a_1}^\dagger{a_1}\ket{vac}=\left(0, 0, \kett{1}{0}, \kett{1}{1} 
      \right),\quad 
    {a_1}{a_2}^\dagger\ket{vac}=\left(0, 0, \kett{0}{1}, * \right),
        \nonumber \\
  && {a_1}^\dagger{a_2}\ket{vac}=\left(0, \kett{1}{0},0, * \right),\quad 
     {a_2}^\dagger{a_2}\ket{vac}=\left(0, \kett{0}{1}, 0, \kett{1}{1} 
      \right). \nonumber
\end{eqnarray}

\noindent Therefore
\begin{eqnarray}
  \label{eq:basematrices}
&&\bra{vac}a_1\ket{vac}=
   \left(
     \begin{array}{ccccc}
       0& 0& 1& 0 \\
       0& 0& 0& 1 \\
       0& 0& 0& 0 \\
       0& 0& 0& 0 
     \end{array}
   \right)
\equiv {\hat{B}}_1\ , \quad
\bra{vac}{a_1}^\dagger\ket{vac}=
   \left(
     \begin{array}{ccccc}
       0& 0& 0& 0 \\
       0& 0& 0& 0 \\
       1& 0& 0& 0 \\
       0& 1& 0& 0 
     \end{array}
   \right)
\equiv {{\hat{B}}_1}^\dagger\ , \nonumber \\
&&\bra{vac}a_2\ket{vac}=
   \left(
     \begin{array}{ccccc}
       0& 1& 0& 0 \\
       0& 0& 0& 0 \\
       0& 0& 0& 1 \\
       0& 0& 0& 0 
     \end{array}
   \right)
\equiv {\hat{B}}_2\ , \quad
\bra{vac}{a_2}^\dagger\ket{vac}=
   \left(
     \begin{array}{ccccc}
       0& 0& 0& 0 \\
       1& 0& 0& 0 \\
       0& 0& 0& 0 \\
       0& 0& 1& 0 
     \end{array}
   \right)
\equiv {{\hat{B}}_2}^\dagger\ , \nonumber \\
&& \bra{vac}{a_1}^2\ket{vac}
   =\bra{vac}\left({a_1}^\dagger\right)^2\ket{vac}
   =\bra{vac}{a_2}^2\ket{vac}
   =\bra{vac}\left({a_2}^\dagger\right)^2\ket{vac}= \hat{\mbox{O}}\ , 
\nonumber \\
&&\bra{vac}{a_1}{a_2}\ket{vac}=
   \left(
     \begin{array}{ccccc}
       0& 0& 0& 1 \\
       0& 0& 0& 0 \\
       0& 0& 0& 0 \\
       0& 0& 0& 0 
     \end{array}
   \right)
=\  {\hat{B}}_1{\hat{B}}_2\ , \
\bra{vac}{a_1}^\dagger{a_2}^\dagger\ket{vac}=
   \left(
     \begin{array}{ccccc}
       0& 0& 0& 0 \\
       0& 0& 0& 0 \\
       0& 0& 0& 0 \\
       1& 0& 0& 0 
     \end{array}
   \right)
=\  {{\hat{B}}_1}^\dagger{{\hat{B}}_2}^\dagger\ , \nonumber \\
&&\bra{vac}{a_1}^\dagger{a_1}\ket{vac}=
   \left(
     \begin{array}{ccccc}
       0& 0& 0& 0 \\
       0& 0& 0& 0 \\
       0& 0& 1& 0 \\
       0& 0& 0& 1 
     \end{array}
   \right)
=\ {{\hat{B}}_1}^\dagger{\hat{B}}_1\ ,\ 
\bra{vac}{a_1}{a_2}^\dagger\ket{vac}=
   \left(
     \begin{array}{ccccc}
       0& 0& 0& 0 \\
       0& 0& 1& 0 \\
       0& 0& 0& 0 \\
       0& 0& 0& 0 
     \end{array}
   \right)
=\ {\hat{B}}_1{{\hat{B}}_2}^\dagger\,  \nonumber \\
&&\bra{vac}{a_1}^\dagger{a_2}\ket{vac}=
   \left(
     \begin{array}{ccccc}
       0& 0& 0& 0 \\
       0& 0& 0& 0 \\
       0& 1& 0& 0 \\
       0& 0& 0& 0 
     \end{array}
   \right)
=\ {{\hat{B}}_1}^\dagger{\hat{B}}_2\ ,\  
\bra{vac}{a_2}^\dagger{a_2}\ket{vac}=
   \left(
     \begin{array}{ccccc}
       0& 0& 0& 0 \\
       0& 1& 0& 0 \\
       0& 0& 0& 0 \\
       0& 0& 0& 1 
     \end{array}
   \right)
=\ {{\hat{B}}_2}^\dagger{\hat{B}}_2\ , \nonumber \\
&&\bra{vac}{\bf 1}\ket{vac}=
   \left(
     \begin{array}{ccccc}
       1& 0& 0& 0 \\
       0& 1& 0& 0 \\
       0& 0& 1& 0 \\
       0& 0& 0& 1 
     \end{array}
   \right)
\equiv\ \hat{E}.
\end{eqnarray}

Under the preceding preliminaries we can determine the connection form 
(\ref{eq:manycoefficients}).  
This is our main result in this paper. \par

\noindent{\bfseries Proposition 3-3-2} 
\begin{eqnarray}
  \label{eq:mainresults}
&&A_{\alpha_1}=\bra{vac}Z^{-1}\frac{\partial}{\partial \alpha_1}Z\ket{vac}
   \nonumber  \\ 
&&= 
 \Bigg\{
   \frac{\bar{\alpha}_1}{2}+ \cosh(\vert\beta_{1}\vert)\bar{c}_0+ 
   \frac{\bar{\beta}_{1}\sinh(\vert\beta_{1}\vert)}{\vert\beta_{1}\vert}c_0
 \Bigg\}\hat{E}  \nonumber  \\ 
 &&+  
 \Bigg\{ 
   \cosh(\vert\beta_{1}\vert)\bar{c}_3+ 
   \frac{\bar{\beta}_{1}\sinh(\vert\beta_{1}\vert)}{\vert\beta_{1}\vert}c_1
 \Bigg\}{\hat{B}}_1+ 
 \Bigg\{ 
   \cosh(\vert\beta_{1}\vert)\bar{c}_1+ 
   \frac{\bar{\beta}_{1}\sinh(\vert\beta_{1}\vert)}{\vert\beta_{1}\vert}c_3
 \Bigg\}{{\hat{B}}_1}^{\dagger}  \nonumber  \\ 
 &&+
 \Bigg\{ 
   \cosh(\vert\beta_{1}\vert)\bar{c}_4+ 
   \frac{\bar{\beta}_{1}\sinh(\vert\beta_{1}\vert)}{\vert\beta_{1}\vert}c_2
 \Bigg\}{\hat{B}}_2+ 
 \Bigg\{ 
   \cosh(\vert\beta_{1}\vert)\bar{c}_2+ 
   \frac{\bar{\beta}_{1}\sinh(\vert\beta_{1}\vert)}{\vert\beta_{1}\vert}c_4
 \Bigg\}{{\hat{B}}_2}^{\dagger}\ ,  \\
 &&{} \nonumber  \\ 
&&A_{\beta_1}=\bra{vac}Z^{-1}\frac{\partial}{\partial \beta_1}Z\ket{vac}
   \nonumber  \\ 
&&= 
 \Bigg\{
   {1\over2}\left(1+ \frac{\sinh(2\beazeta)}{2\beazeta}\right)
   {1\over2}\left({\bar{c}_{0}}^2+ \bar{c}_{1}\bar{c}_{3}+ 
   \bar{c}_{2}\bar{c}_{4}\right)+ 
  \frac{\bar{\beta}_{1}}{2\beazeta^2}\left(-1+\cosh(2\beazeta)\right)
  \nonumber \\
 &&\quad \times  
   {1\over2}\left({\vert c_{0}\vert}^2+ {\vert c_{3}\vert}^2+ 
       {\vert c_{4}\vert}^2+ {1\over2} \right)
 +  
  \frac{{\bar{\beta}_{1}}^2}{2\beazeta^2}
   \left(-1+\frac{\sinh(2\beazeta)}{2\beazeta}\right) 
   {1\over2}\left({c_{0}}^2+ c_{1}c_{3}+ c_{2}c_{4}\right) 
\Bigg\}\hat{E} \nonumber \\
&&+\ 
 \Bigg\{
   {1\over2}\left(1+ \frac{\sinh(2\beazeta)}{2\beazeta}\right)
   \bar{c}_{0}\bar{c}_{3}+ 
  \frac{\bar{\beta}_{1}}{2\beazeta^2}\left(-1+\cosh(2\beazeta)\right)
   {1\over2}\left(c_{1}\bar{c}_{0}+ c_{0}\bar{c}_{3}\right) 
\nonumber \\
&&\quad\ +  
  \frac{{\bar{\beta}_{1}}^2}{2\beazeta^2}
   \left(-1+\frac{\sinh(2\beazeta)}{2\beazeta}\right) 
   c_{0}c_{1} 
\Bigg\}{\hat{B}}_1  \nonumber \\ 
&&+\ 
 \Bigg\{
   {1\over2}\left(1+ \frac{\sinh(2\beazeta)}{2\beazeta}\right)
   \bar{c}_{0}\bar{c}_{1}+ 
  \frac{\bar{\beta}_{1}}{2\beazeta^2}\left(-1+\cosh(2\beazeta)\right)
   {1\over2}\left(c_{3}\bar{c}_{0}+ c_{0}\bar{c}_{1}\right)
\nonumber \\
&&\quad\ +  
  \frac{{\bar{\beta}_{1}}^2}{2\beazeta^2}
   \left(-1+\frac{\sinh(2\beazeta)}{2\beazeta}\right) 
   c_{0}c_{3} 
\Bigg\}{{\hat{B}}_1}^{\dagger} \nonumber \\ 
&&+\ 
 \Bigg\{
   {1\over2}\left(1+ \frac{\sinh(2\beazeta)}{2\beazeta}\right)
   \bar{c}_{0}\bar{c}_{4}+ 
  \frac{\bar{\beta}_{1}}{2\beazeta^2}\left(-1+\cosh(2\beazeta)\right)
   {1\over2}\left(c_{2}\bar{c}_{0}+ c_{0}\bar{c}_{4}\right) 
\nonumber \\
&&\quad\ +  
  \frac{{\bar{\beta}_{1}}^2}{2\beazeta^2}
   \left(-1+\frac{\sinh(2\beazeta)}{2\beazeta}\right) 
   c_{0}c_{2} 
\Bigg\}{\hat{B}}_2  \nonumber \\ 
&&+\ 
 \Bigg\{
   {1\over2}\left(1+ \frac{\sinh(2\beazeta)}{2\beazeta}\right)
   \bar{c}_{0}\bar{c}_{2}+ 
  \frac{\bar{\beta}_{1}}{2\beazeta^2}\left(-1+\cosh(2\beazeta)\right)
   {1\over2}\left(c_{4}\bar{c}_{0}+ c_{0}\bar{c}_{2}\right)
\nonumber \\
&&\quad\ +  
  \frac{{\bar{\beta}_{1}}^2}{2\beazeta^2}
   \left(-1+\frac{\sinh(2\beazeta)}{2\beazeta}\right) 
   c_{0}c_{4} 
\Bigg\}{{\hat{B}}_2}^{\dagger}  \nonumber \\ 
&&+\ 
 \Bigg\{
   {1\over2}\left(1+ \frac{\sinh(2\beazeta)}{2\beazeta}\right)
   \bar{c}_{3}\bar{c}_{4}+ 
  \frac{\bar{\beta}_{1}}{2\beazeta^2}\left(-1+\cosh(2\beazeta)\right)
   {1\over2}\left(c_{2}\bar{c}_{3}+ c_{1}\bar{c}_{4}\right)
\nonumber \\
&&\quad\ +  
  \frac{{\bar{\beta}_{1}}^2}{2\beazeta^2}
   \left(-1+\frac{\sinh(2\beazeta)}{2\beazeta}\right) 
   c_{1}c_{2} 
\Bigg\}{\hat{B}}_1{\hat{B}}_2  \nonumber \\ 
&&+\ 
 \Bigg\{
   {1\over2}\left(1+ \frac{\sinh(2\beazeta)}{2\beazeta}\right)
   \bar{c}_{1}\bar{c}_{2}+ 
  \frac{\bar{\beta}_{1}}{2\beazeta^2}\left(-1+\cosh(2\beazeta)\right)
   {1\over2}\left(c_{4}\bar{c}_{1}+ c_{3}\bar{c}_{2}\right)
\nonumber \\
&&\quad\ +  
  \frac{{\bar{\beta}_{1}}^2}{2\beazeta^2}
   \left(-1+\frac{\sinh(2\beazeta)}{2\beazeta}\right) 
   c_{3}c_{4} 
\Bigg\}{{\hat{B}_1}}^{\dagger}{{\hat{B}}_2}^{\dagger}  \nonumber \\ 
&&+\ 
 \Bigg\{
   {1\over2}\left(1+ \frac{\sinh(2\beazeta)}{2\beazeta}\right)
   \bar{c}_{2}\bar{c}_{3}+ 
  \frac{\bar{\beta}_{1}}{2\beazeta^2}\left(-1+\cosh(2\beazeta)\right)
   {1\over2}\left(c_{4}\bar{c}_{3}+ c_{1}\bar{c}_{2}\right)
\nonumber \\
&&\quad\ +  
  \frac{{\bar{\beta}_{1}}^2}{2\beazeta^2}
   \left(-1+\frac{\sinh(2\beazeta)}{2\beazeta}\right) 
   c_{1}c_{4} 
\Bigg\}{\hat{B}}_1{{\hat{B}}_2}^{\dagger}  \nonumber \\ 
&&+\ 
 \Bigg\{
   {1\over2}\left(1+ \frac{\sinh(2\beazeta)}{2\beazeta}\right)
   \bar{c}_{1}\bar{c}_{4}+ 
  \frac{\bar{\beta}_{1}}{2\beazeta^2}\left(-1+\cosh(2\beazeta)\right)
   {1\over2}\left(c_{2}\bar{c}_{1}+ c_{3}\bar{c}_{4}\right)
\nonumber \\
&&\quad\ +  
  \frac{{\bar{\beta}_{1}}^2}{2\beazeta^2}
   \left(-1+\frac{\sinh(2\beazeta)}{2\beazeta}\right) 
   c_{2}c_{3} 
\Bigg\}{{\hat{B}}_1}^{\dagger}{{\hat{B}}_2}  \nonumber \\ 
&&+\ 
 \Bigg\{
   {1\over2}\left(1+ \frac{\sinh(2\beazeta)}{2\beazeta}\right)
   \bar{c}_{1}\bar{c}_{3}+ 
  \frac{\bar{\beta}_{1}}{2\beazeta^2}\left(-1+\cosh(2\beazeta)\right)
   {1\over2}\left({\vert c_{1}\vert}^2+ {\vert c_{3}\vert}^2\right)
\nonumber \\
&&\quad\ +  
  \frac{{\bar{\beta}_{1}}^2}{2\beazeta^2}
   \left(-1+\frac{\sinh(2\beazeta)}{2\beazeta}\right) 
   c_{1}c_{3} 
\Bigg\}{{\hat{B}}_1}^{\dagger}{{\hat{B}}_1}  \nonumber \\ 
&&+\ 
 \Bigg\{
   {1\over2}\left(1+ \frac{\sinh(2\beazeta)}{2\beazeta}\right)
   \bar{c}_{2}\bar{c}_{4}+ 
  \frac{\bar{\beta}_{1}}{2\beazeta^2}\left(-1+\cosh(2\beazeta)\right)
   {1\over2}\left({\vert c_{2}\vert}^2+ {\vert c_{4}\vert}^2\right)
\nonumber \\
&&\quad\ +  
  \frac{{\bar{\beta}_{1}}^2}{2\beazeta^2}
   \left(-1+\frac{\sinh(2\beazeta)}{2\beazeta}\right) 
   c_{2}c_{4} 
\Bigg\}{{\hat{B}}_2}^{\dagger}{{\hat{B}}_2} , \\
 &&{} \nonumber  \\ 
&&A_{\xi}=\bra{vac}Z^{-1}\frac{\partial}{\partial \xi}Z\ket{vac}
   \nonumber  \\ 
&&= 
 \Bigg\{
   {1\over2}\left(1+ \frac{\sin(2\xizeta)}{2\xizeta}\right)
   \frac{\bar{\zeta}\sinh(2\zezeta)}{2\zezeta}
   \left({\alpha_{2}}^2+ d_1d_2\right)- 
  \frac{\bar{\xi}}{2\xizeta^2}\left(1-\cos(2\xizeta)\right)
  {1\over2}\left({\vert \alpha_{2}\vert}^2+ {\vert d_{2}\vert}^2 \right)
  \nonumber \\ 
&&\quad +  
  \frac{\bar{\xi}^2}{2\xizeta^2}
   \left(-1+\frac{\sin(2\xizeta)}{2\xizeta}\right)
   \frac{\zeta\sinh(2\zezeta)}{2\zezeta} 
   \left({\bar{\alpha_2}}^2+ \bar{d_1}\bar{d_2}\right) 
\Bigg\}\hat{E} \nonumber \\
&&+ 
 \Bigg\{
  \frac{\bar{\xi}^2}{2\xizeta^2}
   \left(-1+\frac{\sin(2\xizeta)}{2\xizeta}\right)
   \cosh(2\zezeta)\bar{\alpha_2}
 \Bigg\}{{\hat{B}}_1} + 
 \Bigg\{
   {1\over2}\left(1+\frac{\sin(2\xizeta)}{2\xizeta}\right)
   \cosh(2\zezeta){\alpha_2}
 \Bigg\}{{\hat{B}}_1}^\dagger  \nonumber \\ 
&&+  
 \Bigg\{
   {1\over2}\left(1+ \frac{\sin(2\xizeta)}{2\xizeta}\right)
   \frac{\bar{\zeta}\sinh(2\zezeta)}{2\zezeta}
   2\alpha_2d_1 - 
  \frac{\bar{\xi}}{2\xizeta^2}\left(1-\cos(2\xizeta)\right)
   {1\over2}\left(\bar{\alpha_2}d_1+ \alpha_2\bar{d_2}\right) 
\nonumber \\ 
&&\quad\ +  
  \frac{\bar{\xi}^2}{2\xizeta^2}
   \left(-1+\frac{\sin(2\xizeta)}{2\xizeta}\right)
   \frac{\zeta\sinh(2\zezeta)}{2\zezeta} 
   2\bar{\alpha_2}\bar{d_2} 
\Bigg\}{{\hat{B}}_2} \nonumber \\
&&+ 
 \Bigg\{
   {1\over2}\left(1+ \frac{\sin(2\xizeta)}{2\xizeta}\right)
   \frac{\bar{\zeta}\sinh(2\zezeta)}{2\zezeta}
   2\alpha_2d_2 - 
  \frac{\bar{\xi}}{2\xizeta^2}\left(1-\cos(2\xizeta)\right)
  {1\over2}\left({\alpha_2}\bar{d_1}+ \bar{\alpha_2}{d_2}\right)
 \nonumber \\  
&&\quad\  +  
  \frac{\bar{\xi}^2}{2\xizeta^2}
   \left(-1+\frac{\sin(2\xizeta)}{2\xizeta}\right)
   \frac{\zeta\sinh(2\zezeta)}{2\zezeta} 
   2\bar{\alpha_2}\bar{d_1} 
\Bigg\}{{\hat{B}}_2}^\dagger \nonumber \\
&&+ 
 \Bigg\{
   {1\over2}\left(1+\frac{\sin(2\xizeta)}{2\xizeta}\right)
   \cosh(2\zezeta)d_1
 \Bigg\}{{\hat{B}}_1}^\dagger{{\hat{B}}_2}+ 
 \Bigg\{
  \frac{\bar{\xi}^2}{2\xizeta^2}
   \left(-1+\frac{\sin(2\xizeta)}{2\xizeta}\right)
   \cosh(2\zezeta)\bar{d_1}
 \Bigg\}{{\hat{B}}_1}{{\hat{B}}_2}^\dagger  \nonumber \\ 
&&+ 
 \Bigg\{
  \frac{\bar{\xi}^2}{2\xizeta^2}
   \left(-1+\frac{\sin(2\xizeta)}{2\xizeta}\right)
   \cosh(2\zezeta)\bar{d_2}
 \Bigg\}{{\hat{B}}_1}{{\hat{B}}_2}+
 \Bigg\{
   {1\over2}\left(1+\frac{\sin(2\xizeta)}{2\xizeta}\right)
   \cosh(2\zezeta)d_2
 \Bigg\}{{\hat{B}}_1}^\dagger{{\hat{B}}_2}^\dagger  \nonumber \\ 
&&+\ 
\frac{\bar{\xi}}{2\xizeta^2}\left(1-\cos(2\xizeta)\right)
{1\over2}{{\hat{B}}_1}^\dagger{{\hat{B}}_1} \nonumber \\ 
&&+\ 
 \Bigg\{
   {1\over2}\left(1+ \frac{\sin(2\xizeta)}{2\xizeta}\right)
   \frac{\bar{\zeta}\sinh(2\zezeta)}{2\zezeta}
   2d_1d_2- 
  \frac{\bar{\xi}}{2\xizeta^2}\left(1-\cos(2\xizeta)\right) 
   {1\over2}\left({\vert d_{1}\vert}^2+ {\vert d_{2}\vert}^2 \right)
  \nonumber \\ 
&&\quad\ +  
  \frac{\bar{\xi}^2}{2\xizeta^2}
   \left(-1+\frac{\sin(2\xizeta)}{2\xizeta}\right)
   \frac{\zeta\sinh(2\zezeta)}{2\zezeta} 
   2\bar{d_1}\bar{d_2}
\Bigg\}{{\hat{B}}_2}^\dagger{{\hat{B}}_2}\ , \\
 &&{} \nonumber  \\ 
&&A_{\zeta}=\bra{vac}Z^{-1}\frac{\partial}{\partial \zeta}Z\ket{vac}
   \nonumber  \\ 
&&=
 \frac{\bar{\zeta}}{2\zezeta^2}(-1+ \cosh(2\zezeta))
   \left(1+ {\vert \alpha_{2}\vert}^2+ {\vert d_{2}\vert}^2 \right)\hat{E}
\nonumber \\
&&+
 \frac{\bar{\zeta}^2}{2\zezeta^2}   
 \left(-1+\frac{\sinh(2\zezeta)}{2\zezeta}\right){\alpha}_2 {\hat{B}}_1 
 +
 {1\over2} \left(1+\frac{\sinh(2\zezeta)}{2\zezeta}\right)
 {\bar{\alpha}}_2 {{\hat{B}}_1}^\dagger 
\nonumber \\
&&+
 \frac{\bar{\zeta}}{2\zezeta^2}(-1+ \cosh(2\zezeta))
  \left(\bar{\alpha_2}d_1+ \alpha_2\bar{d_2}\right) {\hat{B}}_2 
+
 \frac{\bar{\zeta}}{2\zezeta^2}(-1+ \cosh(2\zezeta))
  \left(\bar{\alpha_2}d_2+ \alpha_2\bar{d_1}\right) {{\hat{B}}_2}^\dagger
\nonumber \\
&&+
 \frac{\bar{\zeta}^2}{2\zezeta^2}   
 \left(-1+\frac{\sinh(2\zezeta)}{2\zezeta}\right)d_1 
 {{\hat{B}}_1}{{\hat{B}}_2} 
 +
 {1\over2} \left(1+\frac{\sinh(2\zezeta)}{2\zezeta}\right)
 {\bar d}_1 {{\hat{B}}_1}^\dagger{{\hat{B}}_2}^\dagger 
\nonumber \\
&&+
 \frac{\bar{\zeta}^2}{2\zezeta^2}   
 \left(-1+\frac{\sinh(2\zezeta)}{2\zezeta}\right)d_2 
 {{\hat{B}}_1}{{\hat{B}}_2}^\dagger 
 +
 {1\over2} \left(1+\frac{\sinh(2\zezeta)}{2\zezeta}\right)
 {\bar d}_2 {{\hat{B}}_1}^\dagger{{\hat{B}}_2} 
\nonumber \\
&&+ 
 \frac{\bar{\zeta}}{2\zezeta^2}(-1+ \cosh(2\zezeta))
  {{\hat{B}}_1}^\dagger{{\hat{B}}_1} 
+
 \frac{\bar{\zeta}}{2\zezeta^2}(-1+ \cosh(2\zezeta))
 \left({\vert d_1\vert}^2+ {\vert d_{2}\vert}^2 \right)
  {{\hat{B}}_2}^\dagger{{\hat{B}}_2}\ , \\
 &&{} \nonumber  \\ 
&&A_{\alpha_2}=\bra{vac}Z^{-1}\frac{\partial}{\partial \alpha_2}Z\ket{vac}
=
  {\bar{\alpha}_2\over2}\hat{E}+\cosh\bebzeta {{\hat{B}}_2}^\dagger
  +{\bar{\beta}_2\sinh\bebzeta\over\bebzeta}{{\hat{B}}_2}\ , \\
 &&{} \nonumber  \\ 
&&A_{\beta_2}=\bra{vac}Z^{-1}\frac{\partial}{\partial \beta_2}Z\ket{vac}
=
  {{\bar{\beta}_2(-1+\cosh(2\bebzeta))}\over{2\bebzeta^2}}
  {1\over2}\left({{\hat{B}}_2}^\dagger{{\hat{B}}_2}+{1\over2}\hat{E}\right)\ . 
\end{eqnarray}

\vspace{5mm}
Anyway we determined each term of the connection.  
Since the connection form $\cala$ is anti-hermitian 
($\cala^{\dagger}=-\cala$), it can be written as
\begin{eqnarray}
  \label{eq:calalala}
 &&\cala =
  A_{\alpha_1} d{\alpha_1} + A_{\beta_1} d{\beta_1} +  
  A_\xi d\xi + A_\zeta d\zeta + 
  A_{\alpha_2} d{\alpha_2}+ A_{\beta_2} d{\beta_2} \nonumber \\
 &&\quad - \Bigg\{
    A_{\alpha_1}^{\dagger}d{\bar{\alpha}_1}+ 
    A_{\beta_1}^{\dagger}d{\bar{\beta}_1}+
    {A_\xi}^{\dagger}d{\bar \xi}+{A_\zeta}^{\dagger}d{\bar \zeta}+ 
    A_{\alpha_2}^{\dagger}d{\bar{\alpha}_2}+ 
    A_{\beta_2}^{\dagger}d{\bar{\beta}_2}
   \Bigg\} \ .
\end{eqnarray}
But it is not easy (almost difficult ?) to calculate the 
curvature form $\calf = d\cala + \cala\wedge\cala$ making use of 
Proposition 3-3-2 like Theorem 3-1-4 from Proposition 3-1-3 and 
Theorem 3-2-4 from Proposition 3-2-3.  
This is a miserable task for us.

\section{Problem on Universality}

Here we list our important problem. What should be done in Holonomic 
Quantum Computer is to prove a universality (\cite{JPa}) $\cdots$ a holonomy 
group is irreducible in our terminology, see the below of (\ref{eq:holonomy}).

Let us recall our model in sect. 3.3. The operator $O_1(\alpha_1,\beta_1)$ 
operates 1-st qubit space $\fukuso_{(1)}^2$, $O_2(\alpha_2,\beta_2)$ 
operates 2-nd qubit space $\fukuso_{(2)}^2$ and $W(\xi,\zeta)$ operates 
2-qubit space $\fukuso_{(1)}^2 \otimes \fukuso_{(2)}^2$ intertwiningly. 
Therefore we call 
\begin{equation}
    Z(\alpha_1,\beta_1,\xi,\zeta,\alpha_2,\beta_2)=
    O_1(\alpha_1,\beta_1)W(\xi,\zeta)O_2(\alpha_2,\beta_2)
\end{equation}
one set of fundamental operators. Since the parameter space is 
\begin{equation}
   \calm = \left\{(\alpha_1,\beta_1,\xi,\zeta,\alpha_2,\beta_2 )
            \in \fukuso^6 \right\},
\end{equation}
its dimension is $\mbox{dim}_{\real}\calm$ = $2\ \times\ 6$ = 12,  
while $\mbox{dim}_{\real}U(4) = 4^2$ = 16. 
  The canonical connection on quantum computational bundle on $\calm$ is 
just $\cala$ in (\ref{eq:calalala}).

  Next we operate one set of fundamental operators twice :
\begin{equation}
  Z^{\diamond 2}Z^{\diamond 1}\equiv 
    Z({\tilde \alpha}_1,{\tilde \beta}_1,{\tilde \xi},
    {\tilde \zeta},{\tilde \alpha}_2,{\tilde \beta}_2) 
    Z(\alpha_1,\beta_1,\xi,\zeta,\alpha_2,\beta_2).
\end{equation}
This parameter space is therefore 
\begin{equation}
  {\widetilde \calm} = \left\{({\tilde \alpha}_1,{\tilde \beta}_1,
     {\tilde \xi},{\tilde \zeta},{\tilde \alpha}_2,{\tilde \beta}_2, 
     \alpha_1,\beta_1,\xi,\zeta,\alpha_2,\beta_2 )
     \in \fukuso^{12} \right\},
\end{equation}
its dimension is $\mbox{dim}_{\real}{\widetilde \calm}$ = $2 \times 12$ 
= 24 $\geq$ 16 = $\mbox{dim}_{\real}U(4)$.
 
  We expect that a canonical connection ${\widetilde \cala}$ on quantum 
computational bundle on ${\widetilde \calm}$ like $\cala$ on $\calm$ 
is irreducible., namely $Hol({\widetilde \cala})=U(4)$.

Since each term of curvature form is $F_{\mu \nu}=\partial_{\mu}A_{\nu}-
\partial_{\nu}A_{\mu}+[A_{\mu},A_{\nu}]$, this contains a commutator term 
$[A_{\mu},A_{\nu}]$. From (\ref{eq:basematrices}) we have 
\begin{equation}
  \label{eq:remaining-matrices-1}
 [{\hat{B}}_2, {\hat{B}}_1{{\hat{B}}_2}^\dagger ] = 
   \left(
     \begin{array}{cccc}
       0& 0& 1& 0 \\
       0& 0& 0& -1 \\
       0& 0& 0& 0 \\
       0& 0& 0& 0 
     \end{array}
   \right)\ , \quad 
 [{\hat{B}}_1, {{\hat{B}}_1}^\dagger{\hat{B}}_2 ] = 
   \left(
     \begin{array}{cccc}
       0& 1& 0& 0 \\
       0& 0& 0& 0 \\
       0& 0& 0& -1 \\
       0& 0& 0& 0 
     \end{array}
   \right)\ .
\end{equation}
From these 
\begin{equation}
  \label{eq:remaining-matrices-2}
 [{{\hat{B}}_1}^\dagger{\hat{B}}_2,{{\hat{B}}_2}^\dagger ] = 
   \left(
     \begin{array}{cccc}
       0& 0& 0& 0 \\
       0& 0& 0& 0 \\
       1& 0& 0& 0 \\
       0& -1& 0& 0 
     \end{array}
   \right)\ , \quad 
 [{\hat{B}}_1{{\hat{B}}_2}^\dagger, {{\hat{B}}_1}^\dagger ] = 
   \left(
     \begin{array}{cccc}
       0& 0& 0& 0 \\
       1& 0& 0& 0 \\
       0& 0& 0& 0 \\
       0& 0& -1& 0 
     \end{array}
   \right)\ .
\end{equation}

  Adding (\ref{eq:remaining-matrices-1}) and 
(\ref{eq:remaining-matrices-2}) to (\ref{eq:basematrices}) we have 
\begin{eqnarray}
  \label{eq:remaining-matrices}
    \Bigg\{&& 
     {\hat{E}}, {{\hat{B}}_1}, {{\hat{B}}_1}^\dagger, 
      {{\hat{B}}_2}, {{\hat{B}}_2}^\dagger, 
     {{\hat{B}}_1}{{\hat{B}}_2}, {{\hat{B}}_1}^\dagger{{\hat{B}}_2}, 
     {{\hat{B}}_1}{{\hat{B}}_2}^\dagger, {{\hat{B}}_1}^\dagger{{\hat{B}}_1}, 
     {{\hat{B}}_2}^\dagger{{\hat{B}}_2}, 
     {{\hat{B}}_1}^\dagger{{\hat{B}}_2}^\dagger, \nonumber \\
     &&[{{\hat{B}}_1}, {{\hat{B}}_1}^\dagger{{\hat{B}}_2}], 
     [{{\hat{B}}_2}, {{\hat{B}}_1}{{\hat{B}}_2}^\dagger], 
     [{{\hat{B}}_1}{{\hat{B}}_2}^\dagger, {{\hat{B}}_1}^\dagger], 
     [{{\hat{B}}_1}^\dagger{{\hat{B}}_2}, {{\hat{B}}_2}^\dagger]
    \Bigg\}. 
\end{eqnarray}
They are linearly independent in the Lie algebra $u(4)$ of $U(4)$ and 
$\mbox{dim}_{\real}(\ref{eq:remaining-matrices}) = 15 < 16 = 
\mbox{dim}_{\real}u(4)$. A matrix is lacking, for example,  
\[
   \left(
     \begin{array}{cccc}
       1& 0& 0& 0 \\
       0& -1& 0& 0 \\
       0& 0& -1& 0 \\
       0& 0& 0& 1 
     \end{array}
   \right)\ .
\]
On the other hand it is easy to see
\begin{equation}
  \label{eq:new-matrix}
 [{{\hat{B}}_1}, [{{\hat{B}}_1}^\dagger{{\hat{B}}_2}, {{\hat{B}}_2}^\dagger]] 
 = [{{\hat{B}}_1}, {{\hat{B}}_1}^\dagger] 
   [{{\hat{B}}_2}, {{\hat{B}}_2}^\dagger] 
 = 
   \left(
     \begin{array}{cccc}
       1& 0& 0& 0 \\
       0& -1& 0& 0 \\
       0& 0& -1& 0 \\
       0& 0& 0& 1 
     \end{array}
   \right)\ .
\end{equation}

We note that matrices
\[
   \left(
     \begin{array}{cccc}
       1& 0& 0& 0 \\
       0& 1& 0& 0 \\
       0& 0& 1& 0 \\
       0& 0& 0& 1 
     \end{array}
   \right), \quad 
   \left(
     \begin{array}{cccc}
       0& 0& 0& 0 \\
       0& 0& 0& 0 \\
       0& 0& 1& 0 \\
       0& 0& 0& 1 
     \end{array}
   \right), \quad 
   \left(
     \begin{array}{cccc}
       0& 0& 0& 0 \\
       0& 1& 0& 0 \\
       0& 0& 0& 0 \\
       0& 0& 0& 1 
     \end{array}
   \right), \quad 
   \left(
     \begin{array}{cccc}
       1& 0& 0& 0 \\
       0& -1& 0& 0 \\
       0& 0& -1& 0 \\
       0& 0& 0& 1 
     \end{array}
   \right)
\]
are linearly independent. Therefore the set 
\begin{equation}
    \Bigg\{ (\ref{eq:remaining-matrices}),\  
 [{{\hat{B}}_1},\ [{{\hat{B}}_1}^\dagger{{\hat{B}}_2},{{\hat{B}}_2}^\dagger]] 
    \Bigg\}     
\end{equation}
is linearly independent in $u(4)$ and its dimension is just 16 !
\par 
\vspace{3mm}
\noindent
A comment is in order. To drive a matrix in the RHS of 
(\ref{eq:new-matrix}) in the framework of Holonomic Quantum Computation seems
not easy. How can we get this matrix ?  In my opinion {\textbf two -
dimensional holonomy} are hopeful. For this subject the paper 
\cite{AFG} is recommended. See also \cite{FS2}. 
For more than 3-qubit case a system of generalised 
holonomies $\cdots$ usual holonomy, two-dimensional holonomy, etc $\cdots$ 
should be taken into consideration.
This point will be discussed in the forthcoming paper \cite{KF7}. 
Holonomic Quantum Computation will become more and more complicated. 
\par 
\vspace{10mm}
\noindent
 
Let us introduce an usual approach (a soft one). To prove a universality 
we have only to construct the Controlled--Not operator 
\begin{equation}
  \label{eq:c-not}
\mbox{C-NOT} = 
   \left(
     \begin{array}{ccccc}
       1& 0& 0& 0 \\
       0& 1& 0& 0 \\
       0& 0& 0& 1 \\
       0& 0& 1& 0 
     \end{array}
   \right)
\end{equation}
, see the recent review paper \cite{EHI} or \cite{KF8}. 
Here we define an operator 
\begin{equation}
  \label{eq:an operator}
\mbox{X} = 
   \left(
     \begin{array}{ccccc}
       1& 0& 0& 0 \\
       0& 1& 0& 0 \\
       0& 0& 1& 0 \\
       0& 0& 0& -1 
     \end{array}
   \right).
\end{equation}
For the Hadamard operator $H=
 \frac{1}{\sqrt{2}} 
   \left(
     \begin{array}{cc}
       1& 1 \\
       1& -1 
     \end{array}
   \right)
$ $\left( H^{-1} = H,\  \mbox{or}\  H^2 = E \right)$ . it is easy to see 
\begin{equation}
 \left(E\otimes H\right)X\left(E\otimes H\right)= \mbox{C-NOT}\quad 
 \mbox{or}\quad 
 X= \left(E\otimes H\right)\mbox{C-NOT}\left(E\otimes H\right). 
\end{equation}
Therefore we have only to construct the operator $X$ in Holonomic 
Quantum Computer. To construct $X$ we have only to find a loop
\begin{equation} 
\gamma : [0,1] \longrightarrow 
\fukuso^{6}\equiv \{(\alpha_1,\beta_1,\xi,\zeta,\alpha_2,\beta_2)\} 
\end{equation}
such that 
\begin{equation}
  \Gamma_{\cala}(\gamma) \equiv {\cal P}exp\left\{\oint_{\gamma}\cala\right\} 
  = X  
\end{equation}
for $\cala$ in (\ref{eq:calalala}). To calculate this we use so-called 
non-abelian Stokes theorem, for example, see \cite{BB}. But to use this 
theorem we must calculate the curvature form whether we like it or not.

\section{Further Generalization}

In this section we make a generalization of coherent operators based on 
$\fukuso$ and $su(2)$ and $su(1,1)$ \cite{KF5} and \cite{KF6}, and propose 
a generalization of the method in sect.3.3.

Let us first consider extended coherent operators based on Lie algebras 
$\fukuso$ and $su(1,1)$. 

\noindent{\bfseries Definition}\quad We set 
\begin{eqnarray}
  \label{eq:ed-operator}
  {\rm [NC]}&&\ 
D(\alpha,s) = e^{\alpha a^\dagger - \bar{\alpha}a + isN} \quad 
  {\rm for}\ \   (\alpha,s) \in \fukuso \times \real , \\
  \label{eq:es-operator}
  {\rm [NC]}&&\ 
S(\beta,t) = e^{\beta{\widetilde K}_{+} - \bar{\beta}{\widetilde K}_{-} 
   + 2it{\widetilde K}_{3}}\quad 
  {\rm for}\ \  (\beta,t) \in \fukuso \times \real .
\end{eqnarray}

\noindent Here let us list the disentangling formulas.

\noindent{\bfseries Lemma 5-1}\quad We have 
\begin{eqnarray}
  \label{eq:ed-formula}
{\rm [NC]}\ D(\alpha,s) &=&
  \mbox{exp}\{g(s)\alzeta^2 \}
  \mbox{exp}\{f(s)\alpha a^\dagger \}\mbox{exp}\{isN \}
  \mbox{exp}\{-f(s)\bar{\alpha}a \}, \\
  \mbox{where}&&\ f(s)=\frac{e^{is}-1}{is},\ \ 
                       g(s)=\frac{e^{is}-(1+is)}{s^2},  \nonumber \\
  \label{eq:es-formula}
{\rm [NC]}\ S(\beta,t) &=&
   \mbox{exp}\left\{ \frac{\gamma}{1-ih}{\widetilde K}_{+}\right\}
   \mbox{exp}\left\{ \mbox{log}\frac{1+h^2-\gazeta^2}{(1-ih)^2}
                   {\widetilde K}_{3}\right\}
   \mbox{exp}\left\{ \frac{\bar{\gamma}}{1-ih}{\widetilde K}_{-}\right\}, \\
   \mbox{where}&&\ \tilde{\kappa}=\sqrt{\bezeta^2-t^2},\ 
      \gamma=\frac{\tanh{\tilde{\kappa}}}{\tilde{\kappa}}\beta, \
      \bar{\gamma}=\frac{\tanh{\tilde{\kappa}}}{\tilde{\kappa}}\bar{\beta},\ 
      h=\frac{\tanh{\tilde{\kappa}}}{\tilde{\kappa}}t 
\end{eqnarray}
For the proof see \cite{KF6} and appendix B.

Now we define an operator like $O(\alpha,\beta)$ in (\ref{eq:huni}) :
\begin{equation}
  \label{eq:ext-dsoperator}
  O(\alpha,s,\beta,t)=D(\alpha,s)S(\beta,t). 
\end{equation}
This operator operates on 1-qubit space $\fukuso^2$.

Next let us consider extended coherent operators based on Lie algebras 
$su(2)$ and $su(1,1)$. 

\noindent{\bfseries Definition}\quad We set 
\begin{eqnarray}
  \label{eq:ej-operator}
  {\rm [C]}&&\ 
U(\xi,u) = e^{\xi J_{+} - \bar{\xi}J_{-} + 2iuJ_{3}} \quad 
  {\rm for}\ \   (\xi,u) \in \fukuso \times \real , \\
  \label{eq:ek-operator}
  {\rm [NC]}&&\ 
V(\zeta,v) = e^{\zeta K_{+} - \bar{\zeta}K_{-} + 2ivK_{3}}\quad 
  {\rm for}\ \  (\zeta,v) \in \fukuso \times \real .
\end{eqnarray}

\noindent Here let us list the disentangling formulas.

\noindent{\bfseries Lemma 5-2}\quad We have 
\begin{eqnarray}
  \label{eq:eu-formula}
{\rm [C]}\ U(\xi,u) &=&
  \mbox{exp}\{ \frac{\mu}{1-ik}J_{+} \}
  \mbox{exp}\{ \mbox{log}\frac{1+k^2+\muzeta^2}{(1-ik)^2}J_{3} \}
  \mbox{exp}\{-\frac{\bar\mu}{1-ik}J_{-} \}, \\
   \mbox{where}&&\ \lambda=\sqrt{\xizeta^2+u^2},\ 
      \mu=\frac{\tan{\lambda}}{\lambda}\xi, \
      \bar{\mu}=\frac{\tan{\lambda}}{\lambda}\bar{\xi},\ 
      k=\frac{\tan{\lambda}}{\lambda}u. \nonumber \\ 
{\rm [NC]}\ V(\zeta,v) &=&
   \mbox{exp}\left\{ \frac{\nu}{1-il}K_{+}\right\}
   \mbox{exp}\left\{ \mbox{log}\frac{1+l^2-\nuzeta^2}{(1-il)^2}K_{3}\right\}
   \mbox{exp}\left\{ \frac{\bar{\nu}}{1-il}K_{-}\right\}, \\
   \mbox{where}&&\ \kappa=\sqrt{\zezeta^2-v^2},\ 
      \nu=\frac{\tanh{\kappa}}{\kappa}\zeta, \
      \bar{\nu}=\frac{\tanh{\kappa}}{\kappa}\bar{\zeta},\ 
      l=\frac{\tanh{\kappa}}{\kappa}v. \nonumber 
  \label{eq:ev-formula}
\end{eqnarray}
For the proof see appendix B.

Now we define an operator like $W(\xi,\zeta)$ in (\ref{eq:double}) :
\begin{equation}
  \label{eq:ext-uvoperator}
  W(\xi,u,\zeta,v)=U(\xi,u)V(\zeta,v). 
\end{equation}
This operator operates on 2-qubit space $\fukuso_{(1)}^2 \otimes 
\fukuso_{(2)}^2$. 

In the following we consider a generalization of sect.3.3. First of all 
let us define a full operator
\begin{equation}
  \label{eq:ext-triple}
  \widehat{Z}(\alpha_1,s_1,\beta_1,t_1,\xi,u,\zeta,v,
   \alpha_2,s_2,\beta_2,t_2)= 
     O_{1}(\alpha_1,s_1,\beta_1,t_1)
      W(\xi,u,\zeta,v)
     O_{2}(\alpha_2,s_2,\beta_2,t_2)
\end{equation}
and a family of Hamiltonians
\begin{eqnarray}
  \label{eq:manyfamily}
  &&H_{(\alpha_1,s_1,\beta_1,t_1,\xi,u,\zeta,v,
       \alpha_2,s_2,\beta_2,t_2)}= \nonumber \\
  &&\widehat{Z}(\alpha_1,s_1,\beta_1,t_1,\xi,u,\zeta,v,
  \alpha_2,s_2,\beta_2,t_2)
  H_0 
   \widehat{Z}(\alpha_1,s_1,\beta_1,t_1,\xi,u,\zeta,v,
    \alpha_2,s_2,\beta_2,t_2)^{-1}
\end{eqnarray}
with $H_0$ in (\ref{eq:hamiltonian}). \par 
\noindent We note that our parameter space is 
\begin{equation}
  \widehat{\calm} = 
 \left\{(\alpha_1,s_1,\beta_1,t_1,\xi,u,\zeta,v,\alpha_2,s_2,\beta_2,t_2)
            \in \fukuso^6 \times \real^6 \right\},
\end{equation}
and its dimension is $\mbox{dim}_{\real}\widehat{\calm}$ = $3\ \times\ 6$ 
= 18 $>$ 16 = $\mbox{dim}_{\real}U(4)$. 
 
  We want to calculate a canonical connection ${\widehat \cala}$ on quantum 
computational bundle on ${\widehat \calm}$ like\  ${\cala}$ on ${\calm}$ 
\ (\ref{eq:calalala})\ and to study its holonomy group. 
But this is very hard task.  
We leave it to the readers with brute power of calculations.

\vspace{10mm}
\noindent{\em Acknowledgment.}\quad 
The author wishes to thank Prof. Akio Hosoya who suggested me to perform 
the calculations in sect. 3.3.

\par 
\vspace{10mm}

\begin{Large}
\noindent{\bfseries Appendix A : Some Useful Formulas}
\end{Large}
\par \vspace{5mm} \noindent
Let us recall
\[
   U(\xi)= \mbox{exp}\left(\xi{a_1}^\dagger{a_2}- 
               \bar{\xi}{a_2}^\dagger{a_1}\right),\quad 
   V(\zeta)= \mbox{exp}\left(\zeta{a_1}^\dagger{a_2}^\dagger- 
               \bar{\zeta}{a_2}{a_1}\right). 
\]
\par \noindent 
From this 
\begin{eqnarray}
  \label{eq:adjoint-u}
  {U(\xi)}^{-1}{a_1}U(\xi)&=& \cos(\xizeta)a_1 + 
                          \frac{\xi\sin(\xizeta)}{\xizeta}a_2\ , 
  \nonumber \\
  {U(\xi)}^{-1}{a_1}^{\dagger}U(\xi)&=& \cos(\xizeta){a_1}^\dagger + 
                    \frac{\bar{\xi}\sin(\xizeta)}{\xizeta}{a_2}^\dagger\ , 
  \nonumber \\
  {U(\xi)}^{-1}{a_2}U(\xi)&=& \cos(\xizeta)a_2 - 
                          \frac{\bar{\xi}\sin(\xizeta)}{\xizeta}a_1\ , 
  \nonumber \\
  {U(\xi)}^{-1}{a_2}^{\dagger}U(\xi)&=& \cos(\xizeta){a_2}^{\dagger} -
                      \frac{\xi\sin(\xizeta)}{\xizeta}{a_1}^{\dagger}\ . 
\end{eqnarray}
\par \noindent 
Then 
\begin{eqnarray}
  \label{eq:adjoint-umatrix}
 &&\left({U(\xi)}^{-1}{a_1}U(\xi),{U(\xi)}^{-1}{a_2}U(\xi),
     {U(\xi)}^{-1}{a_1}^{\dagger}U(\xi),{U(\xi)}^{-1}{a_2}^{\dagger}U(\xi)
   \right) \nonumber \\
 &&= \left(a_1,a_2,{a_1}^\dagger,{a_2}^\dagger \right)
   \left(
    \begin{array}{cccc}
     \cos(\xizeta)& -\frac{\bar{\xi}\sin(\xizeta)}{\xizeta}& 0& 0 \\
     \frac{\xi\sin(\xizeta)}{\xizeta}& \cos(\xizeta)&0 &0    \\
     0& 0& \cos(\xizeta)&  -\frac{\xi\sin(\xizeta)}{\xizeta} \\
     0& 0& \frac{\bar{\xi}\sin(\xizeta)}{\xizeta}& \cos(\xizeta)
    \end{array}
   \right)   \nonumber \\
 &&\equiv \left(a_1,a_2,{a_1}^\dagger,{a_2}^\dagger \right){\bf M_U}\ .
\end{eqnarray}

Next 
\begin{eqnarray}
  \label{eq:adjoint-v}
  {V(\zeta)}^{-1}{a_1}V(\zeta)&=& \cosh(\zezeta)a_1 + 
                        \frac{\zeta\sinh(\zezeta)}{\zezeta}{a_2}\dagger\ , 
  \nonumber \\
  {V(\zeta)}^{-1}{a_1}^{\dagger}V(\zeta)&=& \cosh(\zezeta){a_1}^\dagger + 
                 \frac{\bar{\zeta}\sinh(\zezeta)}{\zezeta}{a_2}\ , 
  \nonumber \\
  {V(\zeta)}^{-1}{a_2}V(\zeta)&=& \cosh(\zezeta)a_2 +  
                       \frac{{\zeta}\sinh(\zezeta)}{\zezeta}{a_1}^\dagger\ , 
  \nonumber \\
  {V(\zeta)}^{-1}{a_2}^{\dagger}V(\zeta)&=& \cosh(\zezeta){a_2}^{\dagger} +
                    \frac{\bar{\zeta}\sinh(\zezeta)}{\zezeta}{a_1}\ . 
\end{eqnarray}

Then 
\begin{eqnarray}
  \label{eq:adjoint-vmatrix}
 &&\left({V(\zeta)}^{-1}{a_1}V(\zeta),{V(\zeta)}^{-1}{a_2}V(\zeta),
 {V(\zeta)}^{-1}{a_1}^{\dagger}V(\zeta),{V(\zeta)}^{-1}{a_2}^{\dagger}V(\zeta)
   \right) \nonumber \\
 &&= \left(a_1,a_2,{a_1}^\dagger,{a_2}^\dagger \right)
   \left(
    \begin{array}{cccc}
     \cosh(\zezeta)& 0& 0& \frac{\bar{\zeta}\sinh(\zezeta)}{\zezeta}   \\
     0& \cosh(\zezeta)& \frac{\bar{\zeta}\sinh(\zezeta)}{\zezeta}& 0   \\
     0& \frac{\zeta\sinh(\zezeta)}{\zezeta} & \cosh(\zezeta)& 0     \\
     \frac{\zeta\sinh(\zezeta)}{\zezeta}& 0& 0& \cosh(\zezeta)
    \end{array}
   \right)   \nonumber \\
 &&\equiv \left(a_1,a_2,{a_1}^\dagger,{a_2}^\dagger \right){\bf M_V}\ .
\end{eqnarray}

Here we set 
\begin{eqnarray}
  \label{eq:adjoint-wmatrix}
 &&\left({W(\xi,\zeta)}^{-1}{a_1}W(\xi,\zeta),
   {W(\xi,\zeta)}^{-1}{a_2}W(\xi,\zeta),
   {W(\xi,\zeta)}^{-1}{a_1}^{\dagger}W(\xi,\zeta),
   {W(\xi,\zeta)}^{-1}{a_2}^{\dagger}W(\xi,\zeta)
   \right) \nonumber \\
&&= \left(a_1,a_2,{a_1}^\dagger,{a_2}^\dagger \right){\bf M_W}\ .
\end{eqnarray}

Let us calculate :
\begin{eqnarray}
  \label{eq:adjoint-uvmatrix}
 &&\left({W}^{-1}{a_1}W,{W}^{-1}{a_2}W,{W}^{-1}{a_1}^{\dagger}W,
    {W}^{-1}{a_2}^{\dagger}W \right) 
\nonumber \\
&&= \left({V}^{-1}{U}^{-1}a_1UV,{V}^{-1}{U}^{-1}a_2UV,
     {V}^{-1}{U}^{-1}{a_1}^{\dagger}UV,{V}^{-1}{U}^{-1}{a_2}^{\dagger}UV
   \right) \nonumber \\
&&= {V}^{-1}\left({U}^{-1}a_1U,{U}^{-1}a_2U,
      {U}^{-1}{a_1}^{\dagger}U,{U}^{-1}{a_2}^{\dagger}U \right)V
   \nonumber \\
&&= \left({V}^{-1}a_1V,{V}^{-1}a_2V,
      {V}^{-1}{a_1}^{\dagger}V,{V}^{-1}{a_2}^{\dagger}V \right){\bf M_U}
   \nonumber \\
&&= \left(a_1,a_2,{a_1}^\dagger,{a_2}^\dagger \right){\bf M_V}{\bf M_U}
\end{eqnarray}
Namely, ${\bf M_W} = {\bf M_V}{\bf M_U}$. \quad Let us calculate 
${\bf M_W}$ :
\begin{equation}
  \label{eq:w-matrix}
M_W = 
   \left(
    \begin{array}{cccc}
       \cosh(\zezeta)\cos(\xizeta)& 
       -\cosh(\zezeta)\frac{\bar{\xi}\sin(\xizeta)}{\xizeta}&
       \frac{\bar{\zeta}\sinh(\zezeta)}{\zezeta}
           \frac{\bar{\xi}\sin(\xizeta)}{\xizeta}&
       \frac{\bar{\zeta}\sinh(\zezeta)}{\zezeta}\cos(\xizeta) \\
      \cosh(\zezeta)\frac{{\xi}\sin(\xizeta)}{\xizeta}&
      \cosh(\zezeta)\cos(\xizeta)&
      \frac{\bar{\zeta}\sinh(\zezeta)}{\zezeta}\cos(\xizeta)&
      -\frac{\bar{\zeta}\sinh(\zezeta)}{\zezeta}
           \frac{{\xi}\sin(\xizeta)}{\xizeta} \\
      \frac{{\zeta}\sinh(\zezeta)}{\zezeta} 
           \frac{{\xi}\sin(\xizeta)}{\xizeta}&
      \frac{{\zeta}\sinh(\zezeta)}{\zezeta}\cos(\xizeta)&
      \cosh(\zezeta)\cos(\xizeta)&
      -\cosh(\zezeta)\frac{{\xi}\sin(\xizeta)}{\xizeta} \\
     \frac{{\zeta}\sinh(\zezeta)}{\zezeta}\cos(\xizeta)&
     -\frac{{\zeta}\sinh(\zezeta)}{\zezeta} 
        \frac{\bar{\xi}\sin(\xizeta)}{\xizeta}&
      \cosh(\zezeta)\frac{\bar{\xi}\sin(\xizeta)}{\xizeta}&
      \cosh(\zezeta)\cos(\xizeta)
    \end{array}
   \right)   
\end{equation}

Moreover since
\begin{eqnarray}
  \label{eq:adjoint-o}
  && {O_2}^{-1}{a_1}O_2 = a_1,\quad 
   {O_2}^{-1}{a_1}^{\dagger}O_2 = {a_1}^{\dagger}\ , \nonumber \\
  && {O_2}^{-1}{a_2}O_2 = \alpha_2 + \cosh(\bebzeta)a_2 + 
     \frac{\beta_2\sinh(\bebzeta)}{\bebzeta}{a_2}^\dagger \ , \nonumber \\
  && {O_2}^{-1}{a_2}^{\dagger}O_2 = \bar{\alpha}_2 + 
       \cosh(\bebzeta){a_2}^\dagger + 
       \frac{\bar{\beta}_2\sinh(\bebzeta)}{\bebzeta}{a_2} \ ,
\end{eqnarray}

we have in matrix form 
\begin{eqnarray}
  \label{eq:adjoint-omatrix}
&&\left({\bf 1},{O_2}^{-1}{a_1}O_2,{O_2}^{-1}{a_2}O_2,
     {O_2}^{-1}{a_1}^{\dagger}O_2,{O_2}^{-1}{a_2}^{\dagger}O_2
  \right) \nonumber \\
&&\   = \left({\bf 1},{a_1},{a_2}, {a_1}^\dagger,{a_2}^\dagger \right)
   \left(
    \begin{array}{ccccc}
    1& 0& \alpha_2& 0& \bar{\alpha}_2 \\
    0& 1& 0& 0& 0 \\
    0& 0& \cosh(\bebzeta)& 0& \frac{\bar{\beta}_2\sinh(\bebzeta)}{\bebzeta} \\
    0& 0& 0& 1& 0 \\
    0& 0& \frac{{\beta}_2\sinh(\bebzeta)}{\bebzeta}& 0& \cosh(\bebzeta)
    \end{array}
   \right)  \nonumber \\
&&\  \equiv 
   \left({\bf 1},{a_1},{a_2}, {a_1}^\dagger, {a_2}^\dagger \right)
   {\bf {\widetilde M}_O}.
\end{eqnarray}

Here we define for later convenience 
\begin{equation}
  {\bf {\widetilde M}_W}= 
   \left(
    \begin{array}{cc}
      1& 0 \\
      0& {\bf M_W}
    \end{array}
   \right)\ ,  
\end{equation}

which is in matrix form 
\begin{eqnarray}
  \label{eq:half-matrix}
&&{\bf {\widetilde M}_W}\ =  \nonumber \\
&& 
   \left(
    \begin{array}{ccccc}
       1& 0& 0& 0& 0   \\
       0& 
       \cosh(\zezeta)\cos(\xizeta)& 
       -\cosh(\zezeta)\frac{\bar{\xi}\sin(\xizeta)}{\xizeta}&
       \frac{\bar{\zeta}\sinh(\zezeta)}{\zezeta}
           \frac{\bar{\xi}\sin(\xizeta)}{\xizeta}&
       \frac{\bar{\zeta}\sinh(\zezeta)}{\zezeta}\cos(\xizeta) \\
      0&
      \cosh(\zezeta)\frac{{\xi}\sin(\xizeta)}{\xizeta}&
      \cosh(\zezeta)\cos(\xizeta)&
      \frac{\bar{\zeta}\sinh(\zezeta)}{\zezeta}\cos(\xizeta)&
      -\frac{\bar{\zeta}\sinh(\zezeta)}{\zezeta}
           \frac{{\xi}\sin(\xizeta)}{\xizeta} \\
      0&
      \frac{{\zeta}\sinh(\zezeta)}{\zezeta} 
           \frac{{\xi}\sin(\xizeta)}{\xizeta}&
      \frac{{\zeta}\sinh(\zezeta)}{\zezeta}\cos(\xizeta)&
      \cosh(\zezeta)\cos(\xizeta)&
      -\cosh(\zezeta)\frac{{\xi}\sin(\xizeta)}{\xizeta} \\
      0&
     \frac{{\zeta}\sinh(\zezeta)}{\zezeta}\cos(\xizeta)&
     -\frac{{\zeta}\sinh(\zezeta)}{\zezeta} 
        \frac{\bar{\xi}\sin(\xizeta)}{\xizeta}&
      \cosh(\zezeta)\frac{\bar{\xi}\sin(\xizeta)}{\xizeta}&
      \cosh(\zezeta)\cos(\xizeta)
    \end{array}
   \right)\ .   \nonumber \\
&&{} 
\end{eqnarray}

Therefore
\begin{eqnarray}
&& \left({\bf 1},{O_2}^{-1}{W}^{-1}{a_1}WO_2,{O_2}^{-1}{W}^{-1}{a_2}WO_2,
  {O_2}^{-1}{W}^{-1}{a_1}^{\dagger}WO_2,{O_2}^{-1}{W}^{-1}{a_2}^{\dagger}WO_2
 \right)  \nonumber \\
&&\  = {O_2}^{-1}
 \left({\bf 1},{W}^{-1}{a_1}W,{W}^{-1}{a_2}W,{W}^{-1}{a_1}^{\dagger}W,
    {W}^{-1}{a_2}^{\dagger}W 
 \right)O_2  \nonumber \\
&&\  = 
 \left({\bf 1},{O_2}^{-1}{a_1}O_2,{O_2}^{-1}{a_2}O_2,
  {O_2}^{-1}{a_1}^{\dagger}O_2, {O_2}^{-1}{a_2}^{\dagger}O_2
 \right){\bf {\widetilde M}_W}  \nonumber \\
&&\  =
   \left({\bf 1},{a_1},{a_2},{a_1}^\dagger,{a_2}^\dagger \right)
   {\bf {\widetilde M}_O}{\bf {\widetilde M}_W}.
\end{eqnarray}

We have only to calculate the matrix multiplication 
${\bf {\widetilde M}_O}{\bf {\widetilde M}_W}$. 
\begin{eqnarray}
&& {\bf {\widetilde M}_O}{\bf {\widetilde M}_W}\ = 
   \left(
    \begin{array}{ccccc}
    1& 0& \alpha_2& 0& \bar{\alpha}_2 \\
    0& 1& 0& 0& 0 \\
    0& 0& \cosh(\bebzeta)& 0& \frac{\bar{\beta}_2\sinh(\bebzeta)}{\bebzeta} \\
    0& 0& 0& 1& 0 \\
    0& 0& \frac{{\beta}_2\sinh(\bebzeta)}{\bebzeta}& 0& \cosh(\bebzeta)
    \end{array}
   \right)\ \times  \nonumber \\
&&
   \left(
    \begin{array}{ccccc}
       1& 0& 0& 0& 0   \\
       0& 
       \cosh(\zezeta)\cos(\xizeta)& 
       -\cosh(\zezeta)\frac{\bar{\xi}\sin(\xizeta)}{\xizeta}&
       \frac{\bar{\zeta}\sinh(\zezeta)}{\zezeta}
           \frac{\bar{\xi}\sin(\xizeta)}{\xizeta}&
       \frac{\bar{\zeta}\sinh(\zezeta)}{\zezeta}\cos(\xizeta) \\
      0&
      \cosh(\zezeta)\frac{{\xi}\sin(\xizeta)}{\xizeta}&
      \cosh(\zezeta)\cos(\xizeta)&
      \frac{\bar{\zeta}\sinh(\zezeta)}{\zezeta}\cos(\xizeta)&
      -\frac{\bar{\zeta}\sinh(\zezeta)}{\zezeta}
           \frac{{\xi}\sin(\xizeta)}{\xizeta} \\
      0&
      \frac{{\zeta}\sinh(\zezeta)}{\zezeta} 
           \frac{{\xi}\sin(\xizeta)}{\xizeta}&
      \frac{{\zeta}\sinh(\zezeta)}{\zezeta}\cos(\xizeta)&
      \cosh(\zezeta)\cos(\xizeta)&
      -\cosh(\zezeta)\frac{{\xi}\sin(\xizeta)}{\xizeta} \\
      0&
     \frac{{\zeta}\sinh(\zezeta)}{\zezeta}\cos(\xizeta)&
     -\frac{{\zeta}\sinh(\zezeta)}{\zezeta} 
        \frac{\bar{\xi}\sin(\xizeta)}{\xizeta}&
      \cosh(\zezeta)\frac{\bar{\xi}\sin(\xizeta)}{\xizeta}&
      \cosh(\zezeta)\cos(\xizeta)
    \end{array}
   \right)\ .   \nonumber \\
&&{} 
\end{eqnarray}

\par \vspace{5mm}
\begin{Large}
\noindent{\bfseries Appendix B : Disentangling Formulas}
\end{Large}
\par \vspace{5mm} \noindent
Let us prove the disentangling formulas Lemma 5-2 
for extended coherent operators.
Let $\rho$ be a representation of Lie group $SU(2) \subset SL(2,\fukuso)$ 
\begin{equation}
      \rho : SL(2,\fukuso) \longrightarrow U(\calh \otimes \calh)
\end{equation}
with some special conditions, see \cite{FKSF1} and \cite{FKSF2}, and
\begin{equation}
      J_{+}=d\rho(j_{+}),\quad J_{-}=d\rho(j_{-}), \quad J_{3}=d\rho(j_{-})
\end{equation}
where
\begin{equation}
     j_{+}=
   \left(
     \begin{array}{cc}
        0& 1\\
        0& 0
     \end{array}
   \right), \quad
     j_{-}=
   \left(
     \begin{array}{cc}
        0& 0\\
        1& 0
     \end{array}
   \right), \quad
     j_{3}=\frac{1}{2}
   \left(
     \begin{array}{cc}
        1& 0\\
        0& -1
     \end{array}
   \right).
\end{equation}
In this case 
\begin{eqnarray}
   &&\mbox{exp}\left(\xi J_{+} - \bar{\xi}J_{-} + 2iu J_{3} \right)
         \nonumber \\
   &&= \mbox{exp}\left(d\rho 
        \left(
           \begin{array}{cc}
                iu        & \xi \\
                -\bar{\xi}& -iu
           \end{array}
        \right)
                \right)
    =\rho \left(
         \mbox{exp} 
        \left(
           \begin{array}{cc}
                iu        & \xi \\
                -\bar{\xi}& -iu
           \end{array}
        \right)
          \right)
    \equiv \rho(\mbox{e}^{A}) .       
\end{eqnarray}
Then it is easy to see
\[
  A^2 = -\left(u^2 + \xizeta^2\right)E \equiv -\lambda^2 E  
\]
, so we have 
\begin{eqnarray}
  \mbox{e}^{A}=\cos\lambda E + \frac{\sin\lambda}{\lambda}A
             = 
        \left(
           \begin{array}{cc}
               \cos\lambda + \frac{\sin\lambda}{\lambda}(iu)& 
                       \frac{\sin\lambda}{\lambda}\xi \\
              - \frac{\sin\lambda}{\lambda}\bar{\xi}& 
                \cos\lambda - \frac{\sin\lambda}{\lambda}(iu)
           \end{array}
        \right).
\end{eqnarray}
For 
$\mbox{e}^{A}=
        \left(
           \begin{array}{cc}
             a & b \\
             c & d
           \end{array}
        \right)\ 
(ad - bc = 1)$, 
the Gauss decomposition of this matrix is given by
\begin{equation}
        \left(
           \begin{array}{cc}
             a & b \\
             c & d
           \end{array}
        \right)
=
        \left(
           \begin{array}{cc}
             1 & \frac{b}{d} \\
             0 & 1
           \end{array}
        \right)\ 
        \left(
           \begin{array}{cc}
             \frac{1}{d} & 0 \\
             0 & d
           \end{array}
        \right)\ 
        \left(
           \begin{array}{cc}
             1 & 0 \\
             \frac{c}{d} & 1
           \end{array}
        \right).
\end{equation}
Since $\rho$ is a representation of Lie group (not Lie algebra) we have 
\begin{eqnarray}
 &&\rho\left(
        \left(
           \begin{array}{cc}
             1 & \frac{b}{d} \\
             0 & 1
           \end{array}
        \right)\ 
        \left(
           \begin{array}{cc}
             \frac{1}{d} & 0 \\
             0 & d
           \end{array}
        \right)\ 
        \left(
           \begin{array}{cc}
             1 & 0 \\
             \frac{c}{d} & 1
           \end{array}
        \right)
   \right)  \nonumber \\
 &&=
 \rho\left(
        \left(
           \begin{array}{cc}
             1 & \frac{b}{d} \\
             0 & 1
           \end{array}
        \right)
   \right)\
 \rho\left(
        \left(
           \begin{array}{cc}
             \frac{1}{d} &  \\
             0 & d
           \end{array}
        \right)
   \right)\
 \rho\left(
        \left(
           \begin{array}{cc}
             1 & 0 \\
             \frac{c}{d} & 1
           \end{array}
        \right)
   \right)   \nonumber \\
&&=
 \rho\left(\mbox{exp}{
        \left(
           \begin{array}{cc}
             0 & \frac{b}{d} \\
             0 & 0
           \end{array}
        \right)}
   \right)\
 \rho\left(\mbox{exp}{
        \left(
           \begin{array}{cc}
             -\mbox{log}d& 0 \\
             0 & \mbox{log}d
           \end{array}
        \right)}
   \right)\
 \rho\left(\mbox{exp}{
        \left(
           \begin{array}{cc}
             0 & 0 \\
             \frac{c}{d} & 0
           \end{array}
        \right)}
   \right) \nonumber \\
&&=
\mbox{exp}{\left(
    d\rho
        \left(
           \begin{array}{cc}
             0 & \frac{b}{d} \\
             0 & 0
           \end{array}
        \right)
      \right)}\ 
\mbox{exp}{\left(
    d\rho
        \left(
           \begin{array}{cc}
             -\mbox{log}d& 0 \\
             0 & \mbox{log}d
           \end{array}
        \right)
      \right)}\ 
\mbox{exp}{\left(
    d\rho
        \left(
           \begin{array}{cc}
             0 & 0 \\
             \frac{c}{d} & 0
           \end{array}
        \right)
      \right)} \nonumber \\
&&=
\mbox{exp}{\left(\frac{b}{d}
    d\rho(j_{+})\right)}\ 
\mbox{exp}{\left(-2\mbox{log}
    d\rho(j_{3})\right)}\ 
\mbox{exp}{\left(\frac{c}{d}
    d\rho(j_{-})\right)} \nonumber \\
&&=
\mbox{exp}\left(\frac{b}{d}J_{+}\right)\
\mbox{exp}\left(-2\mbox{log}dJ_{3}\right)\
\mbox{exp}\left(\frac{c}{d}J_{-}\right) \nonumber \\
&&=
\mbox{exp}\left(\frac{b}{d}J_{+}\right)\
\mbox{exp}\left(\mbox{log}\left(\frac{1}{d^2}\right)J_{3}\right)\
\mbox{exp}\left(\frac{c}{d}J_{-}\right) 
\end{eqnarray}
where 
\begin{eqnarray}
 &&\frac{b}{d}=\frac{\frac{\sin\lambda}{\lambda} \xi}
                  {\cos\lambda-iu\frac{\sin\lambda}{\lambda}}
            = \frac{\frac{\tan\lambda}{\lambda} \xi}
                  {1-iu\frac{\tan\lambda}{\lambda}}  \nonumber \\
 &&\frac{c}{d}=\frac{-\frac{\tan\lambda}{\lambda} \bar{\xi}}
                  {1-iu\frac{\tan\lambda}{\lambda}}  \nonumber \\
 &&\frac{1}{d^2}=\frac{1}
                 {\left(\cos\lambda-iu\frac{\sin\lambda}{\lambda}\right)^2}
   =\frac{\frac{1}{\cos\lambda^2}}
        {\left(1-iu\frac{\tan\lambda}{\lambda}\right)^2}
   =\frac{1+\tan\lambda^2}
        {\left(1-iu\frac{\tan\lambda}{\lambda}\right)^2}.
\end{eqnarray}
For simplicity we set
\begin{equation}
   \mu=\frac{\tan\lambda}{\lambda}\xi, \quad
   \bar{\mu}=\frac{\tan\lambda}{\lambda}\bar{\xi}, \quad
   k=\frac{\tan\lambda}{\lambda}u, 
\end{equation}
then it is easy to see
\begin{equation}
  \frac{b}{d}=\frac{\mu}{1-ik},\quad 
  \frac{c}{d}=\frac{-\bar{\mu}}{1-ik}, \quad
  \frac{1}{d^2}=\frac{1+\muzeta^2+k^2}{(1-ik)^2}.
\end{equation}
We could prove [C] in Lemma 5-2 under some conditions. To remove these 
conditions (to extend from a representation of Lie groups to a representation 
of Lie algebras) we needs some tricks. But we omit the details, see \cite{FS}.

 Similar method is still valid for a representation of 
Lie group $SU(1,1)$ to prove [NC] in Lemma 5-2. But we don't repeat here, so  
leave it to the readers.


\begin{thebibliography}{99}
\bibitem{PS}P. W. Shor : 
\newblock Polynomial-time algorithms for prime factorization and discrete 
logarithms on a quantum computer,
\newblock SIAM J. Computing., 26(1997), 1484,
\newblock quant-ph/9508027.
%
\bibitem{AS} A. Steane : 
\newblock Quantum Computing,
\newblock Rept. Prog. Phys., 61(1998), 117. 
%
\bibitem{RP} E. Rieffel and W. Polak : 
\newblock An Introduction to Quantum Computing for Non-Physicists,
\newblock quant-ph/9809016.
%
\bibitem{LPS}H. K. Lo, S. Popescu and T. Spiller (Eds) : 
\newblock Introduction to quantum computation and information,
\newblock  World Scientific, Singapore, 1999. 
%
\bibitem{JVEC}J. Jones, V. Vedral. A. Ekert and G. Castagnoli : 
\newblock Geometric Quantum Computation with NMR,
\newblock quant-ph/9910052.
%
\bibitem{ZR}P. Zanardi and M. Rasetti : 
\newblock Holonomic Quantum Computation,
\newblock Phys. Lett. A264(1999), 94,
\newblock quant-ph/9904011.
%
\bibitem{PZR}J. Pachos, P. Zanardi and M. Rasetti : 
\newblock Non-Abelian Berry connections for quantum computation,
\newblock to appear in Phys. Rev. A, 
\newblock quant-ph/9907103.
%
\bibitem{AYK}A. Yu. Kitaev : 
\newblock Fault-tolerant quantum computation by anyons,
\newblock quant-ph/9707021. 
%
\bibitem{JP}J. Preskill : 
\newblock Fault-Tolerant Quantum Computation,
\newblock quant-ph/97120408. 
%
\bibitem{SW}A. Shapere and F. Wilczek (Eds) : 
\newblock Geometric Phases in Physics,
\newblock World Scientific, Singapore, 1989.
%
\bibitem{PC}J. Pachos and S. Chountasis : 
\newblock Optical Holonomic Quantum Computer,
\newblock quant-ph/9912093.
%
\bibitem{KF2} K. Fujii : 
\newblock Note on Coherent States and Adiabatic Connections, Curvatures,
\newblock J. Math. Phys.,  
\newblock 41(2000), 4406.
%
\bibitem{KF3} K. Fujii : 
\newblock Mathematical Foundations of Holonomic Quantum Computer,
\newblock quant-ph/0004102.
%
\bibitem{KF4} K. Fujii : 
\newblock More on Optical Holonomic Quantum Computer,
\newblock quant-ph/0005129.
%
\bibitem{MN}M. Nakahara : 
\newblock Geometry, Topology and Physics,
\newblock IOP Publishing Ltd, 1990.
%
\bibitem{KF1}K. Fujii : 
\newblock Solutions of $A_{\infty}$ Toda equations based on noncompact
group $SU(1,1)$ and infinite-dimensional Grassmann manifolds,
\newblock J. Math. Phys., 36(1995), 1652.
%
\bibitem{FKSF1}K. Funahashi, T. Kashiwa, S. Sakoda and K. Fujii : 
\newblock Coherent states, path integral, and semiclassical approximation,
\newblock  J. Math. Phys., 36(1995), 3232.
%
\bibitem{FKSF2}K. Funahashi, T. Kashiwa, S. Sakoda and K. Fujii : 
\newblock Exactness in the Wentzel-Kramers-Brillouin approximation for 
some homogeneous spaces,
\newblock J. Math. Phys., 36(1995), 4590.
%
\bibitem{FKS}K. Fujii, T. Kashiwa, S. Sakoda :
\newblock Coherent states over Grassmann manifolds and the WKB exactness
in path integral,
\newblock J. Math. Phys., 37(1996), 567.
%
\bibitem{AP}A. Perelomov : 
\newblock Generalized Coherent States and Their Applications,
\newblock Springer--Verlag, 1986.
%
\bibitem{FS}K. Fujii and T. Suzuki : 
\newblock A Universal Disentangling Formula for Coherent States of Perelomov's
 Type,
\newblock hep-th/9907049.
%
\bibitem{SLB}S. Seshadri, S. Lakshmibala and V. Balakrishnan : 
\newblock Geometric phases for generalized squeezed coherent states,
\newblock Phys. Rev. A55(1997), 869,
\newblock quant--ph 9905101.
%
\bibitem{JPa}J. Pachos : %
\newblock a private communication. 
%
\bibitem{PZ}J. Pachos and P. Zanardi : 
\newblock Quantum Holonomies for Quantum Computing,
\newblock quant-ph/0007110. 
%
\bibitem{AFG}O. Alvarez, L. A. Ferreira and J. S. Guillen : 
\newblock A New Approach to Integrable Theories in Any Dimension,
\newblock Nucl. Phys. B529(1998), 689, 
\newblock hep-th/9710147.
%
\bibitem{FS2}K. Fujii and T. Suzuki : 
\newblock Nonlinear Sigma Models in (1+2) Dimensions and an Infinite 
Number of Conserved Currents,
\newblock Lett. Math. Phys. 46(1998), 49, 
\newblock hep-th/9802105.
%
\bibitem{KF7}K. Fujii : 
\newblock A System of Generalized Holonomies and Quantum Computation 
 (a tentative title), 
\newblock in progress.
%
\bibitem{EHI}A. Ekert, P. Hayden and H. Inamori : %
\newblock Basic concepts in quantum computation, 
\newblock quant-ph/0011013.
%
\bibitem{KF8}K. Fujii : 
\newblock A Lecture on Quantum Logic Gates, 
\newblock quant-ph/0101054.
%
\bibitem{BB}B. Broda : 
\newblock Non-Abelian Stokes Theorem in Action,  
\newblock math-ph/0012035.
%
\bibitem{KF5}K. Fujii : 
\newblock Basic Properties of Coherent and Generalized Coherent 
Operators Revisited,
\newblock quant-ph/0009012.
%
\bibitem{KF6}K. Fujii : 
\newblock Note on Extended Coherent Operators and Some Basic Properties, 
\newblock quant-ph/0009116.
%
\end{thebibliography}
\end{document}